\documentclass[12pt,preprint]{aastex}

\usepackage{graphicx}
\usepackage{amsmath}
\usepackage{natbib}
\usepackage{bm}

\newcommand{\pan}{{\tt Pandurata}\, }
\newcommand{\harm}{{\tt Harm3D}\, }
\newcommand{\pand}{{\tt Pandurata}}

\newcommand{\grmhd}{{\tt GRMHD}\, }
\newcommand{\grmhdd}{{\tt GRMHD}}

\shorttitle{Monte Carlo Radiation Transport}
\shortauthors{Schnittman \& Krolik}

\begin{document}

\title{A Monte Carlo Code for Relativistic Radiation Transport Around
  Kerr Black Holes}

\author{Jeremy D.\ Schnittman}
\affil{NASA Goddard Space Flight Center \\
Greenbelt, MD 20771}
\email{jeremy.schnittman@nasa.gov}
\and
\author{Julian H.\ Krolik}
\affil{Department of Physics and Astronomy,
Johns Hopkins University\\
Baltimore, MD 21218}
\email{jhk@pha.jhu.edu}

\begin{abstract}
We present a new code for radiation transport around Kerr black holes,
including arbitrary emission and absorption mechanisms, as well as electron
scattering and polarization. The code is particularly useful for
analyzing accretion flows made up of optically thick disks and
optically thin coronae. We give a detailed description of the methods
employed in the code, and also present results from a number of
numerical tests to assess its accuracy and convergence. 
\end{abstract}


\section{INTRODUCTION}\label{section:intro}

Since the beginning of X-ray astronomy over 50 years ago, there has
been steadily growing interest in relativistic radiation
transport. Because of the high energies of both photons and electrons
relevant to these astrophysical sources, special relativistic effects
must be included in most particle interactions. And because the central
engines of so many X-ray sources are compact objects such as neutron
stars and black holes, full general relativity must also be
incorporated into any physically realistic code. 

Here we present in detail for the first time the fully relativistic
Monte Carlo radiation transport code \pand\footnote{The name {\tt
    Pandurata} comes from {\it Coelogyne pandurata}, a species of black
  orchid native to Borneo. Much of
  the core radiation transport is derived from the code {\tt
    Buttercup}, developed by Schnittman for inertial fusion
  applications \citep{schnittman:96,schnittman:00} at the University
  of Rochester's Laboratory for Laser
  Energetics (LLE). In holding with LLE's long tradition of naming
  codes after flowers, the name {\tt Pandurata} was chosen to
  represent the joint heritage of black holes and laboratory radiation
  hydrodynamics}. While this is the first formal description of the
code in the literature, it has been under development for many years
and has already been used in numerous publications
\citep{schnittman:09,schnittman:10,noble:11,schnittman:12}. \pan
shares features with numerous existing codes in the literature, but we
believe its combination of full general relativity, wide range of
emission and absorption processes, scattering, polarization, and
optically thin/thick capabilities make it uniquely valuable in the
rapidly evolving field of black hole astrophysics. Most recently, in
\citet{schnittman:12} we have demonstrated how the code may be used to
take a major step towards bridging the
gap between global magneto-hydrodynamics (MHD) simulations and real
X-ray observations of accreting black holes. 

The literature of radiation transport in astrophysics is extremely
broad and includes scores of different techniques and applications. It
would be a futile endeavor to attempt to give a comprehensive
summary of work here. Rather, we will simply highlight a few recent
contributions that are most relevant to the applications of interest,
namely photon transport around Kerr black holes. By far the most
common approach has been ray-tracing geodesic paths backwards from a
distant observer to the accretion flow, calculating a transfer
function of some sort, and coupling this to some model for emission to
generate spectra and light curves. A few of the many examples of this
observer-to-emitter approach include
\citet{rauch:94,broderick:03,broderick:04,schnittman:04,dovciak:04,schnittman:06,noble:07,dexter:09,dexter:09b}.

A smaller number of codes have been written with the
emitter-to-observer approach, which may be more physically intuitive,
but is almost always more computationally intensive. With the exception of
\citet{laor:90,laor:91,kojima:91}, which use uniform sampling of
emission angles, these codes are
generally Monte Carlo in nature, such as {\tt grmonty} by
\citet{dolence:09}, and the present work. As we will show below,
particularly when electron scattering is included, the
emitter-to-observer paradigm is almost essential for capturing the
most relevant physics of the problem. 

Another feature that is relatively uncommon in these ray-tracing
codes, but of increasing interest in the high energy community, is
polarization. It is included in
\citet{agol:00,dovciak:08,huang:09,shcherbakov:11,huang:11,marin:12}, 
although often only for vacuum transport and not including
scattering. Disk polarization is treated by
\citet{laor:90,matt:93,dovciak:08} for both Schwarzschild and Kerr
black holes, but neglecting electron scattering. \citet{dovciak:11}
includes illumination from a source above the disk, while
\citet{dovciak:08} includes a cold plane-parallel atmosphere above the
disk, geometrically thin with varying optical depth. 
A small number of ray-tracing codes also allow for non-standard black
hole metrics as a way of testing general
relativity. \citet{krawczynski:12} follows the emitter-to-observer
paradigm for calculating polarized flux from a thermal disk, and
\citet{psaltis:12} describes an observer-to-emitter framework that can
be applied to a large number of space-time tests such as timing,
spectra, and imaging
\citep{johannsen:10a,johannsen:10b,johannsen:11,johannsen:12}. 

The body of literature including detailed scattering and polarization
is generally restricted to flat spacetime, and often only the most
simple geometries
\citep{connors:77,connors:80,sunyaev:85,haardt:93,haardt:94,poutanen:96}.
Here we attempt to synthesize the strengths of all these various codes
into a single flexible radiation transport tool for analyzing both
global MHD simulations, and also simpler toy accretion models. The
ultimate goal is to produce concrete predictions that can be compared
directly with the large and continually growing body of X-ray
observations of accreting black holes. 

Precisely because of the large interest in this topic, we give here a
comprehensive description of the technical components of our radiation
transport code \pand. We hope that the techniques outlined below will
be valuable to others who are interested in developing similar (or
even better, more powerful) tools. We also present the results from a
suite of simple numerical tests to verify the code, thus lending
support and increasing our confidence in earlier work based on \pand.

\section{LOCAL ORTHONORMAL FRAMES}\label{section:tetrads}
The most general input for \pan is a body of tabulated data including
the extrinsic fluid variables density, temperature, magnetic
field, and the 4-velocity at each point in a three-dimensional
volume. Multiple data slices in the time coordinate allow for studies
of variability. The coordinates are Boyer-Lindquist for a Kerr BH with
mass $M$ and dimensionless spin parameter $a/M$. The fluid variables
are given in physical cgs units for a specific BH mass and
accretion rate. The source of the data is quite general, and \pan has
been used successfully to analyse simulation data from the
relativistic MHD codes \harm \citep{noble:11, schnittman:12} and \grmhd
\citep{schnittman:06}, as well as 2D hydro simulations
\citep{schnittman:06a} and analytic models for the
accretion disk and corona \citep{schnittman:09,schnittman:10}.

We adopt a $(-+++)$ metric signature, and convention where Greek
indices run from 0 to 3, and Latin indices are restricted to spatial
coordinates 1 to 3.
The coordinate metric is given by \citep{boyer:67}:
\begin{equation}
g_{\mu \nu} = \begin{pmatrix}
-\alpha^2+\omega^2 \varpi^2 & 0 & 0 & -\omega\varpi^2 \\
0 & \rho^2/\Delta & 0 & 0 \\
0 & 0 & \rho^2 & 0 \\
-\omega\varpi^2 & 0 & 0 & \varpi^2 \end{pmatrix}  \, .
\end{equation}

This allows for a relatively simple form for the inverse metric:
\begin{equation}
g^{\mu \nu} = \begin{pmatrix}
-1/\alpha^2 & 0 & 0 & -\omega/\alpha^2 \\
0 & \Delta/\rho^2 & 0 & 0 \\
0 & 0 & 1/\rho^2 & 0 \\
-\omega/\alpha^2 & 0 & 0 & 1/\varpi^2-\omega^2/\alpha^2 \end{pmatrix}  \, .
\end{equation}

In geometrized units with $G=c=1$, we have
\begin{subequations}
\begin{eqnarray}\label{eqn:BL_equations_a}
\rho^2 & \equiv & r^2+a^2\cos^2\theta \\ 
\Delta & \equiv & r^2-2Mr+a^2 \\
\alpha^2 & \equiv & \frac{\rho^2\Delta}{\rho^2\Delta+2Mr(a^2+r^2)} \\
\omega & \equiv & \frac{2Mra}{\rho^2\Delta+2Mr(a^2+r^2)} \\
\varpi^2 & \equiv &
\left[\frac{\rho^2\Delta+2Mr(a^2+r^2)}{\rho^2}\right]\sin^2\theta \label{eqn:BL_equations_e}\, .
\end{eqnarray}
\end{subequations}

\subsection{Simulation Data}\label{section:simulation}
\begin{figure}
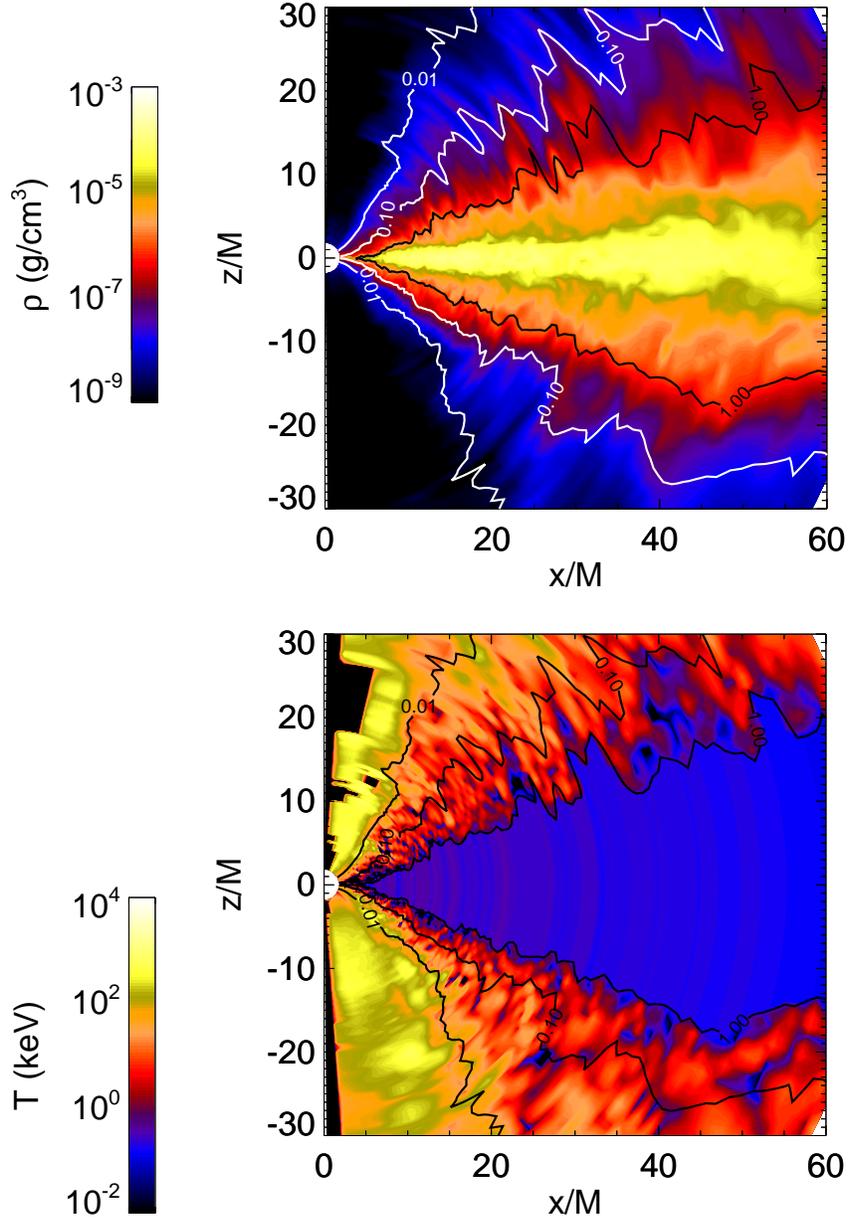

\caption{\label{fig:rho_contour} Fluid density ({\it top}) and
  temperature ({\it bottom}) profile for a slice of \harm
  data in the $(r,z)$ plane, taken from the ThinHR simulation
  \citep{noble:10,noble:11}. Contours show surfaces of constant
  optical depth with $\tau=0.01, 0.1, 1.0$. Fiducial values for the
  black hole mass $M=10M_\odot$ and luminosity $L=0.1L_{\rm Edd}$ were
  used, as described in \citet{schnittman:12}.}
\begin{center}
\scalebox{0.7}{\includegraphics*{harm_rhobar.epsi}}
\hspace{0.5cm}
\scalebox{0.7}{\includegraphics*{harm_rho.epsi}} \\
\vspace{0.5cm}
\scalebox{0.7}{\includegraphics*{harm_Tebar.epsi}}
\hspace{0.5cm}
\scalebox{0.7}{\includegraphics*{harm_Te.epsi}} \\
\end{center}
\end{figure}

We briefly describe here the format of data from the \harm
MHD simulations. Similar data can be generated from \grmhdd, and 
any analytic model can be understood as a subset of the full
tabulated simulation data. As described in greater detail in
\citet{noble:11} and \citet{schnittman:12}, the first step in
post-processing the simulation data is to convert from code units of
density and local dissipation to cgs units of density and
temperature. Given the density everywhere, we integrate
the optical depth along paths of constant
$(r,\phi)$ coordinates starting from both $\theta=0$ and $\theta=\pi$
to get $\tau_{\rm top}(r,\theta,\phi)$ and $\tau_{\rm
  bot}(r,\theta,\phi)$. The disk midplane can then be defined as the
surface $\theta_{\rm mid}(r,\phi)$ where $\tau_{\rm top}(r,\theta_{\rm
  mid},\phi) = \tau_{\rm bot}(r,\theta_{\rm mid},\phi)$. When
$\tau(r,\theta_{\rm mid},\phi) > 1$, the disk is optically thick and
we define a top and bottom photosphere $\Theta(r,\phi)$ such that
$\tau_{\rm top}(r,\Theta_{\rm top},\phi) = \tau_{\rm
  bot}(r,\Theta_{\rm bot},\phi)=1$. In Figure \ref{fig:rho_contour} we
show a slice in the $(r,z)$ plane of simulation data from the \harm
``ThinHR'' run \citep{noble:10}. The local temperature is represented by the
logarithmic color scale, and the contours show surfaces of constant
$\tau$. In Figure \ref{fig:surface_tetrads} we show a
three-dimensional representation of the photosphere surface
$\Theta_{\rm top}(r,\phi)$ for the same simulation data. 

From the photosphere surfaces, thermal photon are launched into the
optically thin corona above and below the disk. Because the opacity
within the disk is
usually dominated by electron scattering, the seed photons are emitted
with the limb-darkening and polarization dependence on angle given by
\citet{chandra:60}. The spectrum is that of a diluted blackbody with
temperature $T_{\rm eff}$ and hardening factor $f$:
\begin{equation}\label{eqn:seed_spectrum}
I_\nu = \frac{1}{f^4}B_\nu(f\, T_{\rm eff}),
\end{equation}
with $B_\nu$ the black-body function. We take $f=1.8$
\citep{shimura:95} and the local effective temperature is given by 
\begin{equation}\label{T_eff}
T_{\rm eff}(r,\phi) \equiv \left(\frac{\mathcal{F}(r,\phi)}{\sigma}\right)^{1/4},
\end{equation}
where $2\mathcal{F}(r,\phi)$ is the total integrated luminosity in the
optically thick part of the disk (the factor of 2 comes from the fact
that the flux is emitted equally from the top and bottom
photospheres). As shown in Figure \ref{fig:rho_contour}, the gas has a
constant temperature inside the disk for a given $(r,\phi)$, due to
the high level of thermalization caused by the large optical depth.

Synchrotron and bremsstrahlung seed photons can also be generated in
the coronal regions, in which case we use an unpolarized, isotropic distribution
function for the emission angles, as measured in the local fluid
frame. Due to the high temperatures and low densities of the
coronal regions, the net power in the coronal seed photons is typically much
lower than that of inverse-Compton scattering from the thermal seeds
coming from the disk \citep{schnittman:12}.

\subsection{Photosphere tetrads}\label{section:phototetrad}
We begin with a short discussion of notation. 
As stressed in \citet{mtw:73}, vectors are
invariant geometric objects independent of coordinate system, and we
represent them with bold font $\mathbf{u}$, while the components in a
specific basis are represented with italics: $u^\mu$. We
adopt a naming convention such that the components of a vector in the
coordinate basis are represented by $\mu$ and in the local fluid
frame by $\hat{\mu}$. 
The basis vectors themselves are labeled with
$(\mu)$. For example, the coordinate basis is spanned by the vectors
$\mathbf{e}_{(\mu)}$ with components $e_{(\mu)}^{\nu} =
\delta_{(\mu)}^{\nu}$, where $\delta$ is the usual Kronneker
delta. Note that the coordinate basis vectors are not normalized, and
not even orthogonal in the Kerr metric:
\begin{equation}
\mathbf{e}_{(\mu)} \cdot \mathbf{e}_{(\nu)} = 
g_{\alpha \beta} e_{(\mu)}^{\alpha} e_{(\nu)}^{\beta} =
g_{\mu \nu} \, .
\end{equation}

Einstein's Equivalence Principle, one of the bedrocks of general
relativity, states that an orthonomal basis (a ``tetrad'') can be
defined at any point in space. In fact, an arbitrary number of tetrads
can be defined at any point, and are all related by Lorentz boosts
and/or rotations. One particularly useful tetrad in the Kerr metric is
that of the zero-angular-momentum observer (ZAMO;
\citet{bardeen:72}). We denote the ZAMO frame with $\tilde{\mu}$
labels. It can be constructed from the coordinate basis by:
\begin{subequations}
\begin{eqnarray}
\mathbf{e}_{(\tilde{t})} &=& \frac{1}{\alpha}\mathbf{e}_{(t)} + 
\frac{\omega}{\alpha}\mathbf{e}_{(\phi)} \\
\mathbf{e}_{(\tilde{r})} &=& \sqrt{\frac{\Delta}{\rho^2}}\mathbf{e}_{(r)} \\
\mathbf{e}_{(\tilde{\theta})} &=& \sqrt{\frac{1}{\rho^2}}\mathbf{e}_{(\theta)} \\
\mathbf{e}_{(\tilde{\phi})} &=&
\sqrt{\frac{1}{\varpi^2}}\mathbf{e}_{(\phi)} \, .
\end{eqnarray}
\end{subequations}
Any vector can be represented by its components in different
bases:
\begin{equation}
\mathbf{u} = \mathbf{e}_{(\mu)}u^\mu =
\mathbf{e}_{(\tilde{\mu})}u^{\tilde{\mu}} \, ,
\end{equation}
and the components are related by a linear transformation
$E_{\,\, \tilde{\mu}}^{\mu}$: 
\begin{subequations}
\begin{eqnarray}
u^\mu &=&  E_{\,\, \tilde{\mu}}^\mu u^{\tilde{\mu}} \, , \\
u^{\tilde{\mu}} &=&  [E^{-1}]_{\,\, \mu}^{\tilde{\mu}} u^{\mu} \, .
\end{eqnarray}
\end{subequations}
In our example of the ZAMO frame, $E_{\,\, \tilde{\mu}}^{\mu}$ is given by
\begin{equation}
E_{\,\, \tilde{\mu}}^{\mu} = \begin{pmatrix}
\frac{1}{\alpha} & 0 & 0 & 0 \\
0 & \sqrt{\frac{\Delta}{\rho^2}} & 0 & 0 \\
0 & 0 & \frac{1}{\rho} & 0 \\
\frac{\omega}{\alpha} & 0 & 0 & \frac{1}{\varpi} 
\end{pmatrix}\, .
\end{equation}

At each point on the photosphere we define a tetrad in the comoving
fluid frame (designated with sub/superscripts $\hat{\mu}$) such that
the time coordinate is in the direction of the fluid 4-velocity:
\begin{equation}
e_{(\hat{t})}^\mu = u^\mu.  
\end{equation}
In our notation, this equation says that the 4-vector tangent to the
world line of an observer moving with the fluid can be expressed in
the Boyer-Lindquist coordinate basis with components $\mu$, or in the
local frame with components $\hat{\mu}$ with $e_{(\hat{t})}^{\hat{\mu}}=[1,0,0,0]$.
The spatial basis vectors in the fluid frame $e_{(\hat{i})}^\mu$ are
constucted via a method similar to 
\citet{beckwith:08}, including a slight modification to ensure the
right-handedness of the basis such that $\mathbf{e}_{(\hat{z})}$ is in the
$-\theta$ direction. For completeness, we reproduce those definitions here:

\begin{subequations}
\begin{eqnarray}
C_0 &=& u_\phi/u_t \, ,\\
C_1 &=& \frac{u^r}{u^t+C_0 u^\phi} \, ,\\
C_2 &=& \frac{g^{tt}+2C_0 g^{t\phi}+C_0^2g^{\phi \phi}}{g^{rr}} \, ,\\
C_3 &=& u^t+C_1 C_2 u^r+C_0 u^\phi \, ,\\
N_1 &=& \sqrt{g_{tt}C_0^2-2g_{t\phi}C_0 +g_{\phi \phi}} \, ,\\
N_2 &=& \sqrt{g^{tt}C_1^2+2g^{t\phi}C_1^2C_0+g^{rr}+g^{\phi\phi}
C_0^2 C_1^2} \, ,\\
N_3 &=& \sqrt{(u^\theta)^2(g^{tt}+2g^{t\phi}C_0+g^{rr}C_1^2 C_2^2+
  g^{\phi \phi} C_0^2)+g^{\theta \theta}C_3^2} \, ,
\end{eqnarray}
\end{subequations}
and
\begin{subequations}
\begin{eqnarray}
e_{(\hat{x})}^\mu &=& \left[-\frac{1}{N_2}(g^{tt}C_1+g^{t\phi}C_0C_1),
\frac{g^{rr}}{N_2}, 0,
-\frac{1}{N_2}(g^{t\phi}C_1+g^{\phi\phi}C_0C_1)\right] \, ,\\
e_{(\hat{y})}^\mu &=&
\left[-\frac{C_0}{N_1},0,0,\frac{1}{N_1}\right] \, ,\\
e_{(\hat{z})}^\mu &=&
\left[\frac{1}{N_3}(g^{tt}u^\theta+g^{t\phi}C_0u^\theta),
\frac{g^{rr}C_1 C_2 u^\theta}{N_3}, -\frac{g^{\theta \theta}C_3}{N_3},
\frac{1}{N_3}(g^{t\phi}u^\theta+g^{\phi\phi}C_0 u^\theta)\right]\, .
\end{eqnarray}
\end{subequations}

From this tetrad basis, any other tetrad in the fluid frame can be
constructed from a simple rotation of the spatial basis vectors. We
take as our preferred basis (now labeled with $\mathbf{e}_{(\bar{\mu})}$) one in
which $\mathbf{e}_{(\bar{z})}$ is normal to the photosphere surface.
Whether we are using simulation data or an analytic model for the disk
surface, the photosphere is described by a two-dimensional
surface on the top and bottom of the disk: $\Theta_{\rm top}(r,\phi)$
and $\Theta_{\rm bot}(r,\phi)$. From these functions, at each point on
the photosphere we can construct two vectors tangent to the disk
surface through the following process. Start with the
coordinate-based vectors
\begin{subequations}
\begin{equation}
dr^\mu = [0,\Delta r,\frac{\partial \Theta}{\partial r} \Delta r, 0]
\end{equation}
and
\begin{equation}
d\phi^\mu = [0,0,\frac{\partial \Theta}{\partial \phi} \Delta\phi,
  \Delta\phi]\, ,
\end{equation}
\end{subequations}
where $\Delta r$ and $\Delta\phi$ are the differential sizes of the
fluid cell in question\footnote{For example, the ThinHR simulation
  uses $\Delta r/r = 0.004$ and $\Delta\phi=\pi/128$.}. 
Next, subtract off the components parallel to $\mathbf{e}_{(\hat{t})}$:
\begin{subequations}
\begin{equation}
\mathbf{dr}' = \mathbf{dr}-(\mathbf{dr}\cdot \mathbf{e}_{(\hat{t})})
\mathbf{e}_{(\hat{t})} 
\end{equation}
and
\begin{equation}
\mathbf{d}\boldsymbol{\phi}' = \mathbf{d}\boldsymbol{\phi}-
\mathbf{d}\boldsymbol{\phi}\cdot \mathbf{e}_{(\hat{t})})
\mathbf{e}_{(\hat{t})} \, .
\end{equation}
\end{subequations}
When $\mathbf{dr}'$ and $\mathbf{d}\bm{\phi}'$ are projected into the fluid
frame, they will have only spatial components and will be tangent to
the photosphere. 
In this basis, we can easily construct the normal
vector by taking the 3-vector cross product:
\begin{equation}\label{eqn:dzdA}
dz^{\hat{k}} = \epsilon_{\,\, \hat{i}\hat{j}}^{\hat{k}} dr'^{\hat{i}}d\phi'^{\hat{j}}\, .
\end{equation}
This procedure has the added advantage of giving the proper area of the
photosphere patch subtended by the vectors $\mathbf{dr}'$ and
$\mathbf{d}\boldsymbol{\phi}'$ by $dA=|\mathbf{dz}|$. This formula for
$dA$ will be helpful for
determining the amplitude of emitted flux from each patch of the
disk, since the emission function is typically defined in the local
fluid frame. Because $\mathbf{dr}'$ and
$\mathbf{d}\boldsymbol{\phi}'$ are not generally orthogonal, we also
define the $\mathbf{dx}$ and $\mathbf{dy}$ tangent vectors by
$\mathbf{dx}=\mathbf{dr}'$, $dy^{\hat{t}}=0$ and 
\begin{equation}
dy^{\hat{k}} = \epsilon_{\,\, \hat{i}\hat{j}}^{\hat{k}} dz^{\hat{i}}dx^{\hat{j}}\, .
\end{equation}

\begin{figure}[t]
\caption{\label{fig:surface_tetrads} Three-dimensional representation
  of the disk photosphere surface $\Theta_{\rm top}(r,\phi)$, along with the local spatial tetrad
  definitions of equation (\ref{eqn:surface_tetrads}). The simulation
  data is the same as in Figure \ref{fig:rho_contour}. The color scale
  is a linear representation of the disk's local thermal
  temperature. The labels $(r_i, \phi_j)$ correspond to coordinates of
  the computational grid boundaries.}
\begin{center}
\scalebox{0.6}{\includegraphics*{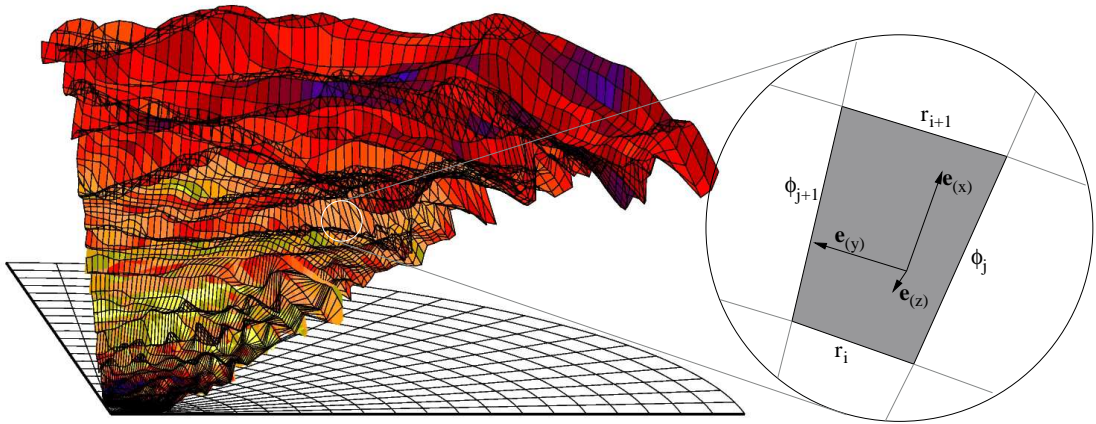}}
\end{center}
\end{figure}

To complete the tetrad, we simply need to normalize the
differential basis vectors. Returning to the coordinate basis, we
have: 
\begin{subequations}
\begin{eqnarray}\label{eqn:surface_tetrads}
e^\mu_{(\bar{t})} &=& e^\mu_{(\hat{t})} = u^\mu \\
e^\mu_{(\bar{x})} &=& dx^\mu/(g_{\alpha \beta}dx^\alpha
dx^\beta)^{1/2} \\
e^\mu_{(\bar{y})} &=& \pm dy^\mu/(g_{\alpha \beta}dy^\alpha
dy^\beta)^{1/2} \\
e^\mu_{(\bar{z})} &=& \pm dz^\mu/(g_{\alpha \beta}dz^\alpha
dz^\beta)^{1/2} \, .
\end{eqnarray}
\end{subequations}
The $\pm$ in the definitions for $\mathbf{e}_{(\bar{y})}$ and
$\mathbf{e}_{(\bar{z})}$ are chosen for the top $(+)$ and bottom $(-)$
photosphere surfaces so that the spatial basis vectors are oriented in
a right-hand fashion and to ensure that $\mathbf{e}_{(\bar{z})}$ points
{\it away} from the disk body. In Figure \ref{fig:surface_tetrads} we
show how these tetrad basis vectors are oriented on the photosphere
surface. 

\subsection{Coronal tetrads}\label{section:coronatetrad}
In addition to launching photons from the disk surface, we often
want the option of including seeds from within the corona, due to
thermal bremsstrahlung, cyclo-synchrotron, or other radiation
processes. Analogous with the comoving surface element defined above
for disk emission, for coronal emission we need to define a volume
element and associated tetrad at each point in the simulation
space. Like the tetrads defined above, the time axis is defined along
the local fluid 4-velocity $u^\mu$. However, unlike the surface tetrads, the
volume tetrads have no preferred orientation\footnote{For some
  specialized emission models, such as optically thin synchrotron, it
  may be convenient to choose a special orientation, e.g., with the
  $\mathbf{e}_{(z)}$ basis rotated to lie along the local
  magnetic field vector.}, so we can simply use the
spatial coordinate vectors projected onto the space orthogonal to
$u^\mu$:
\begin{subequations}
\begin{eqnarray}
dr^\mu &=& [0,\Delta r,0, 0]\, , \\
d\theta^\mu &=& [0,0,\Delta\theta, 0]\, , \\
d\phi^\mu &=& [0,0,0,\Delta\phi]\, ,
\end{eqnarray}
\end{subequations}
\begin{subequations}
\begin{eqnarray}
\mathbf{dr}' &=& \mathbf{dr}-(\mathbf{dr}\cdot
\mathbf{e}_{(\hat{t})})\, , \\
\mathbf{d}\boldsymbol{\theta}' &=& \mathbf{d}\boldsymbol{\theta}
-(\mathbf{d}\boldsymbol{\theta}\cdot
\mathbf{e}_{(\hat{t})})\, , \\
\mathbf{d}\boldsymbol{\phi}' &=& \mathbf{d}\boldsymbol{\phi}-
(\mathbf{d}\boldsymbol{\phi}\cdot \mathbf{e}_{(\hat{t})})\, . \\
\end{eqnarray}
\end{subequations}
The proper volume element subtended by these vectors is given by the
3-vector triple product in the local fluid frame. While there is no
real preferred orientation for the spatial axes, we still need to go
through the process of defining {\it some} orthonormal basis to
project the above vectors and thereby calculate vector
products. In practice, we set $\mathbf{e}_{(\hat{x})}$ along the
$\mathbf{dr}'$ direction:
\begin{subequations}
\begin{equation}
dx^\mu = dr'^\mu ,
\end{equation}
then set the y-axis roughly along the $\phi$ coordinate direction
\begin{equation}
dy^{\hat{k}} = \epsilon_{\,\, \hat{i}\hat{j}}^{\hat{k}} dx^{'\hat{i}}d\theta^{\hat{j}}\, ,
\end{equation}
and the z-axis normal to both:
\begin{equation}
dz^{\hat{k}} = \epsilon_{\,\, \hat{i}\hat{j}}^{\hat{k}} dx^{'\hat{i}}dy^{\hat{j}}\, .
\end{equation}
\end{subequations}
As above for the photosphere tetrads, the final step is to normalize
all the basis vectors:
\begin{subequations}
\begin{eqnarray}
e^\mu_{(\bar{t})} &=& e^\mu_{(\hat{t})} = u^\mu \\
e^\mu_{(\bar{x})} &=& dx^\mu/(g_{\alpha \beta}dx^\alpha
dx^\beta)^{1/2} \\
e^\mu_{(\bar{y})} &=& dy^\mu/(g_{\alpha \beta}dy^\alpha
dy^\beta)^{1/2} \\
e^\mu_{(\bar{z})} &=& dz^\mu/(g_{\alpha \beta}dz^\alpha
dz^\beta)^{1/2} \, .
\end{eqnarray}
\end{subequations}
Unlike the photosphere case, since there is no ``top'' or ``bottom''
in the corona, we need not be concerned about the orientation of the
$\mathbf{e}_{(\bar{z})}$ vector, and simply require a right-handed
(x,y,z) convention.

\section{RAY-TRACING}\label{section:raytracing} 

\subsection{Geodesics}
The ray-tracing portion of \pan integrates the geodesic trajectories
of photons in the Kerr metric. From the tetrad frames defined in the
previous section, the initial direction of a photon is selected from
an isotropic distribution in the emitting fluid frame (limited to a
hemisphere in the case of an optically thick photosphere surface). 

The geodesic integrater is the same as that described in
\citet{schnittman:04}, based on a Hamiltonian approach. Because the
Kerr metric is stationary, the momentum conjugate to the time
coordinate $t$ is conserved, and corresponds to the (negative)
specific energy of a particle ($m^2=1$) or photon ($m^2=0$). We can replace the affine
parameter with the coordinate time and write the Hamiltonian as 
\begin{equation}
H(x^i,p_i) \equiv -p_0 = \frac{g^{0i}p_i}{g^{00}} +
\left[\frac{g^{ij}p_ip_j + m^2}{-g^{00}}
+\left(\frac{g^{0i}p_i}{g^{00}}\right)^2 \right]^{1/2} \, ,
\end{equation}
with equations of motion
\begin{subequations}
\begin{equation}\label{hameq_1}
\frac{dx^i}{dt} = \frac{\partial H_1}{\partial p_i}\, ,
\end{equation}
\begin{equation}\label{hameq_2}
\frac{dp_i}{dt} = -\frac{\partial H_1}{\partial x^i}\, .
\end{equation}
\end{subequations}

In Boyer-Lindquist coordinates, the Hamiltonian can be written thus:
\begin{equation}
H(r,\theta,\phi,p_r,p_\theta,p_\phi) = \omega p_\phi
+\alpha\left(\frac{\Delta}{\rho^2}p_r^2 
+\frac{1}{\rho^2}p_\theta^2 +\frac{1}{\varpi^2} p_\phi^2
+m^2\right)^{1/2}\, ,
\end{equation}
using the same notation defined above in equations
(\ref{eqn:BL_equations_a}-\ref{eqn:BL_equations_e}). 
Because the metric, and thus Hamiltonian, is axisymmetric, $p_\phi$ is
also an integral of the motion. We are thus left with five coupled
first-order ordinary differential
equations for $(r, \theta, \phi, p_r,p_\theta)$. The third integral of
motion, Carter's constant \citep{carter:68}
\begin{equation}
\mathcal{Q} \equiv p_\theta^2
+\cos^2\theta \left[a^2(m^2-p_0^2)+p_\phi^2/\sin^2\theta\right],
\end{equation}
is used as an independent check of the accuracy of the numerical
integration. 

\begin{figure}[t]
\caption{\label{fig:cash_karp} Convergence of the geodesic integrator,
  as determined by the accuracy of the conserved quantity
  $\mathcal{Q}$. As expected, we find 5$^{th}$-order convergence for
  the Cash-Karp adaptive time step integrator.}
\begin{center}
\scalebox{0.6}{\includegraphics*{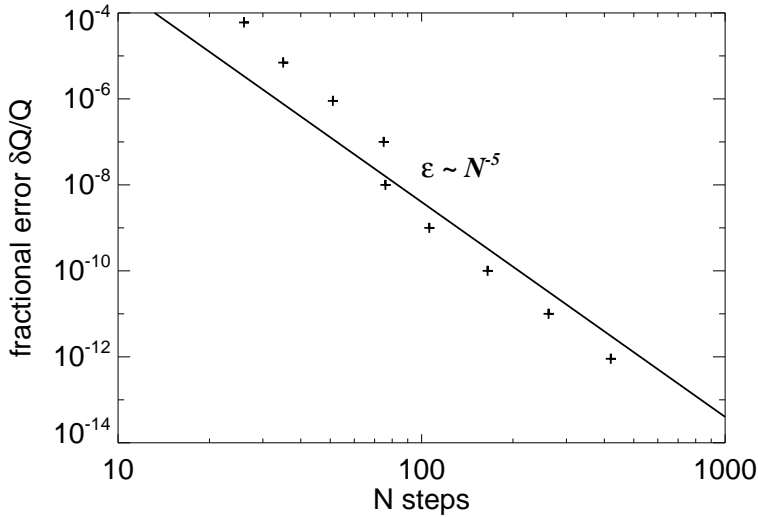}}
\end{center}
\end{figure}

For the numerical integration of geodesics, we use a 5$^{th}$-order
Cash-Karp algorithm with adaptive step size \citep{press:92}. In Figure
\ref{fig:cash_karp} we show the accuracy of the integrator by plotting
the average deviation in the Carter constant for a selection of
photons around a black hole with $a/M=0.99$, as a function of step
segments. We typically set the tolerance at $10^{-8}$ per step, which
we find allows sufficient sampling of the fluid variables near the
black hole. Because of the frequent table look-ups required when using 
simulation data, there is little to be gained by using more advanced
integration techniques such as Bulirsch-Stoer or the semi-analytic
approaches of \citet{rauch:94} or \citet{dexter:09} that calculate the
geodesic endpoint in a single integral evaluation, and are more
appropriate for vacuum transport.

\subsection{Polarization}

\pan is also capable of polarized transport along geodesics. The
polarization vector is a space-like vector normal to the photon
direction. For a photon with wavevector $\mathbf{k}$, the polarization
vector $\mathbf{f}$ is constrained by $\mathbf{f}\cdot\mathbf{f}=1$
and $\mathbf{f}\cdot\mathbf{k}=0$ \citep{connors:80}. The vector
$\mathbf{f}$ is parallel transported along the geodesic path:
$\nabla_\mathbf{k}\mathbf{f}=0$, but instead of explicitly solving the
parallel transport equation, we can take advantage of the
complex-valued Walker-Penrose constant $\kappa_{\rm wp}$
\citep{walker:70, connors:77}. 

After solving for the wavevector $k^\mu$ along the geodesic path, $\kappa_{\rm
  wp}$ is given at any point by
\begin{eqnarray}
\kappa_{\rm wp} &=&
\left[ (k^t f^r - k^r f^t)+a \sin^2\theta (k^r f^\phi - k^\phi f^r) \right.
  \nonumber\\
& & \left. -i[(r^2+a^2)(k^\phi f^\theta- f^\phi k^\theta)-
    a(k^t f^\theta - k^\theta f^t)] \sin\theta \right]
(r-ia \cos\theta) \, .
\end{eqnarray}
Combined with the normalization factors $\mathbf{f}\cdot\mathbf{f}=1$
and $\mathbf{f}\cdot\mathbf{k}=0$, we have four linear equations for
the four components of $f^\mu$. Because $\mathbf{k}$ is a null vector,
we can always redefine $\mathbf{f}$ by a multiple of $\mathbf{k}$:
$\mathbf{f}'=\mathbf{f}+\lambda\mathbf{k}$, and thus write the
polarization vector as 
\begin{equation}
f^\mu = [0, \cos\psi e^i_1+ \sin\psi e^i_2]
\end{equation}
for some space-like basis vectors $\mathbf{e}_1$ and $\mathbf{e}_2$ normal to
$\mathbf{k}$. 

The degree of
polarization $\delta \le 1$ is invariant along the ray path. When
interacting with a distant detector or scattering off an electron in
the fluid frame, it is convenient to employ the classical Stokes
parameters $I$, $Q$, and $U$. In the ($\mathbf{e}_1$, $\mathbf{e}_2$)
basis, we can write
\begin{subequations}
\begin{eqnarray}
X &=& Q/I = \delta \cos 2\psi \, ,\\
Y &=& U/I = \delta \sin 2\psi \, .
\end{eqnarray}
\end{subequations}
One of the main advantages of this approach is that the Stokes
parameters for each photon can simply be added at the detector, quite
useful in a Monte Carlo calculation. Furthermore, $I(\nu)$, $Q(\nu)$,
and $U(\nu)$ can all be written in units of spectral density, which is
the standard observable for many real detectors. 

For photons emitted at an angle $\theta_{\rm em}$ to the normal of a
scattering-dominated surface, we use the results of \citet{chandra:60}
to get the initial polarization amplitude [ranging from
$\delta(\theta_{\rm em}=0^\circ)=0$ up to $\delta(\theta_{\rm
    em}=90^\circ)=0.12$] and direction (parallel to the disk surface). 

\subsection{Photon packets}

Because the geodesic photon trajectories are independent of photon
energy, we can significantly improve the efficiency of the Monte Carlo
calculation by tracking large numbers of photons simultaneously,
covering a range of energies. We call these computational entities
``photon packets,'' which are analogous to the ``superphotons'' of
\citet{dolence:09}, except for the fact that theirs are monoenergetic
and ours are broad-band. We also assign a single polarization angle
and degree to the entire photon packet. This is an approximation that
works well for vaccuum transport and coherent scattering, but will
break down when including 
scattering at high energies $h\nu \gtrsim m_e c^2$ as the electron
cross section becomes more energy-dependent. 

Each photon packet is weighted by a number of geometric emission
factors. For example, a photon packet emitted from a small patch of
optically thick, scattering-dominated accretion disk would have a
spectrum of 
\begin{equation}
F_\nu^{\rm em} = \frac{1}{f^4}B_\nu(f\, T_{\rm eff})\, \frac{1}{u^t}\, f_{\rm
  limb}(\theta_{\rm em})\, \cos\theta_{\rm em} \, dA\, d\Omega \, ,
\end{equation}
where $F_\nu$ is a function that has units of spectral luminosity [erg/s/Hz].
Here $f$ is the same hardening function introduced above in equation
(\ref{eqn:seed_spectrum}), $\cos\theta_{\rm em}$ is a geometric factor for
emission from an optically thick
surface, $f_{\rm limb}$ is a limb-darkening function given by
\citet{chandra:60}, $dA$ is the proper area of the emission region
[see eqn.\ (\ref{eqn:dzdA}) above], and
$d\Omega=2\pi/N_{\rm ph}$ is the proper solid angle of a hemisphere
sampled evenly by $N_{\rm ph}$ photon packets. Lastly,
$1/u^t=d\tau/dt$ is a relativistic correction 
factor to convert from time in the emission frame to that in the
coordinate or distant observer frame. 

In order to account for the spectral redshift, we store both $F$ and
$\nu$ at a set of discrete points. When transforming from the emitter
to observer frames, $F$ is invariant (units of s$^{-1}$ and Hz$^{-1}$
cancel)\footnote{For a discrete function $F_i$, the number of photons
  emitted per {\it coordinate-frame} second between $\nu_i$ and
  $\nu_i+d\nu_i$ is $F_i\, d\nu_i/(h\nu_i)$, where $h$ is Planck's
  constant and $\nu_i$ are measured in the local emission frame. Because
  $\nu_i$ and $d\nu_i$ transform the same under Lorentz
  transformations, $F_i$ is invariant.}, 
while $\nu$ transforms as follows. If the
photon packet is emitted in a frame with fluid 4-velocity $u^\mu({\rm em})$
and photon 4-momentum $k_\mu({\rm em})$,
and observed in a frame with $u^\nu({\rm obs})$ and $k_\nu({\rm obs})$,
then we can write the redshifted frequencies as
\begin{equation}\label{eqn:nu_obs}
\nu({\rm obs}) = \nu({\rm em}) \frac{u^\nu({\rm obs}) k_\nu({\rm obs})}
{u^\mu({\rm em}) k_\mu({\rm em})} \, .
\end{equation}
Whenever the photon packet scatters off the disk or an electron in
the corona, the frequencies $\nu_i$ are updated and the old ``observed''
frame becomes the new ``emitted'' frame.
When the photon packet reaches an observer at infinity, $u^\nu({\rm
  obs})=[1,0,0,0]$ and the well-known redshift relation is
reproduced. 

For this distant observer, the angle of
polarization $\psi$ is measured by projecting $\mathbf{f}$ onto the
$(\mathbf{e}_1,\mathbf{e}_2)=(\mathbf{e}_\phi,-\mathbf{e}_\theta)$
basis. For an observer oriented with the black hole spin axis
projected in the vertical direction, $\psi=0$ corresponds to
horizontal polarization \citep{schnittman:09,schnittman:10}. Given $\psi$, $\delta$, and $F_\nu$, the
spectral luminosity form of the Stokes parameters are simply
\begin{subequations}\label{eqn:QUtilde}
\begin{eqnarray}
\tilde{Q}_\nu &=& F_\nu \delta \cos 2\psi \, ,\\
\tilde{U}_\nu &=& F_\nu \delta \sin 2\psi \, ,
\end{eqnarray}
\end{subequations}
where $\tilde{Q}$ and $\tilde{U}$ are related to the Stokes parameters
$Q$ and $U$ by a factor of $(F/I)$. 
After summing over a large number of photon packets, we then invert
equation (\ref{eqn:QUtilde}) and return to the $\delta(\nu)$,
$\psi(\nu)$ representation.  

\subsection{Emission and absorption}
Along each geodesic path, we can also include local emission and
absorption processes such as bremsstrahlung or synchrotron. This is
the predominant method for generating
light curves and spectra in codes that shoot photons backwards from a
distant observer
\citep{broderick:04,schnittman:04,schnittman:06,noble:07,noble:09,dexter:09}.
In the fluid frame, the radiation transport equation is given by 
\begin{equation}\label{eqn:rad_trans}
\frac{dI_\nu}{ds} = j_\nu -\alpha_\nu I_\nu\ ,
\end{equation}
where $ds$ is the differential path length and $I_\nu$, $j_\nu$, and
$\alpha_\nu$ are respectively the spectral intensity, emissivity, and
absorption coefficient of the local fluid.
The absorption coefficient is related to the opacity
$\kappa_\nu$ through the density $\rho$: $\alpha_\nu =
\rho\kappa_\nu$. Defining the optical depth $\tau_\nu$ through
\begin{equation}\label{eqn:tau_nu}
d\tau_\nu \equiv \alpha_\nu ds,
\end{equation}
the transfer equation can be written as
\begin{equation}\label{eqn:rad_trans_eq2}
\frac{dI_\nu}{d\tau_\nu} = S_\nu - I_\nu,
\end{equation}
where the source function is defined as $S_\nu \equiv
 j_\nu/\alpha_\nu$.

Both $I_\nu$ and $S_\nu$ have the same properties under Lorentz
transformations, namely $I_\nu/\nu^3$ and $S_\nu/\nu^3$ are both
invariant. Other invariant terms are the optical depth $\tau_\nu$,
$\nu\alpha_\nu$, and $j_\nu/\nu^2$ \citep{rybicki:04}. 
Thus if we can solve the non-relativistic radiative transfer equation
(\ref{eqn:rad_trans}) in the local fluid frame, then in any other
inertial frame (e.g., the ZAMO tetrad), the special relativistic
version can be written
\begin{equation}\label{eqn:rad_trans_eq3}
\frac{dI_\nu}{ds} = \left(\frac{\nu}{\nu'}\right)^2 j_\nu' -
\left(\frac{\nu'}{\nu}\right) \alpha_\nu' I_\nu.
\end{equation}
Here the fluid frame (where $j_\nu$ and $\alpha_\nu$ are defined) is
the primed frame, and the ``lab frame'' unprimed. 

The above analysis, while quite useful for special relativistic flows
in the locally flat ZAMO basis, ignores all general relativistic effects of curved
spacetime around the black hole. To include these effects, we need
only shift the frequencies $\nu_i$ from one geodesic step to the next,
due solely to the 
gravitational redshift, and we can treat each computational step as a
new observer relative to the previous step, as in equation
(\ref{eqn:nu_obs}).

\section{SCATTERING}\label{section:scattering}

We allow for two types of scattering in \pand: Compton scattering off
free electrons in the corona, and scattering off an optically thick
disk (which in turn is characterized by repeated scatterings in the
relatively cool atmosphere). Because electron conserves
photon number, our photon packet approach is ideal for modeling
these processes. 

\subsection{Coronal Scattering}\label{section:coronal_scattering}
The first step in the scattering process is to determine whether a
scattering event takes place at all. To do this, we transform into a
local inertial ``lab'' frame, generally taken to be the ZAMO frame
discussed above in Section \ref{section:phototetrad}. In this frame,
the photon moves a distance of $dl^2 =
\eta_{\tilde{i}\tilde{j}}dx^{\tilde{i}} dx^{\tilde{j}}$ in a single geodesic
integration step $dt$. Then the total optical depth to scattering
along the path is 
\begin{equation}
d\tau = dl\, \kappa_{\rm es}\, \rho_{\rm lab} = dl\, \kappa_{\rm es}\,
\rho_{\rm fluid} \frac{\nu_{\rm fluid}}{\nu_{\rm lab}}\, ,
\end{equation}
where the last equality comes from the invariance of $\nu \alpha_\nu$
\citep{rybicki:04}, with the absoption coefficient $\alpha_\nu =
\kappa_{\rm es}\rho$ for electron scattering opacity. Given $d\tau$
(typically much less than unity), the probability of scattering is
$P=1-e^{-d\tau}$.  

When a photon does scatter off a free electron, we carry out the
scattering calculation in the electron's rest frame. This requires two
coordinate transformations: from the coordinate basis (denoted with
$\mu$ super/subscripts) to a fluid-frame
tetrad ($\bar{\mu}$), and then a Lorentz boost from the fluid frame to the
electron's rest frame ($\bar{\mu}'$). The transformation from coordinate basis to
corona fluid frame is the same as given above in Section
\ref{section:coronatetrad}. In the fluid frame, the electron velocity
$\beta = v/c$
is taken from an isotropic Maxwell-Juttner distribution 
\begin{equation}\label{eqn:maxwell_juttner}
f(\gamma) = \frac{\gamma^2 \beta}{\theta_T\, K_2(1/\theta_T)}
\exp(-\gamma/\theta_T)\, ,
\end{equation}
where $\gamma = 1/\sqrt{1-\beta^2}$, $\theta_T = kT/m_ec^2$, and $K_2$
is the modified Bessel function. See Appendix \ref{section:app_juttner}
for a description of our algorithm for generating a Monte Carlo sample
of velocities that satisfy equation (\ref{eqn:maxwell_juttner}).

Following \citet{mtw:73}, we construct a generic Lorentz boost in the
direction of the electron 4-velocity $u^{\bar{\mu}}=[\gamma,\gamma\beta
  n^{\bar{j}}]$:
\begin{eqnarray}
u^{\bar{\mu}} &=& [\gamma,\gamma\beta n^{\bar{j}}] \hspace{1cm} (|n| =
1), \nonumber\\ 
\Lambda^{\bar{t}'}_{\,\, \bar{t}} &=& \gamma, \nonumber\\
\Lambda^{\bar{t}'}_{\,\, \bar{j}} &=& \Lambda^{\bar{j}'}_{\,\, \bar{t}} 
= -\beta \gamma n^{\bar{j}}, \nonumber\\
\Lambda^{\bar{j}'}_{\,\, \bar{k}} &=& \Lambda^{\bar{k}'}_{\,\, \bar{j}} 
= (\gamma-1)n^{\bar{j}} n^{\bar{k}}+\delta^{\bar{j} \bar{k}}.
\end{eqnarray}
The photon momentum in the electron frame is thus given by
$p^{\bar{\mu}'}=\Lambda^{\bar{\mu}'}_{\,\, \bar{\mu}}p^{\bar{\mu}}$.  

\begin{figure}[t]
\caption{\label{fig:scat_schem} Schematic of the scattering geometry
  in the electron frame. The incoming radiation is polarized along the 
  $\bm{\varepsilon}_1$ direction. The scattered radiation
  $\mathbf{k}_f$ makes an angle $\Theta$ with $\bm{\varepsilon}_1$
  and $\theta$ with $\mathbf{k}_i$. When projected onto the
  $\bm{\varepsilon}_1-\bm{\varepsilon}_2$ plane, $\mathbf{k}_f$ makes an
  angle $\phi$ with $\bm{\varepsilon}_1$.}
\begin{center}
\scalebox{0.6}{\includegraphics*{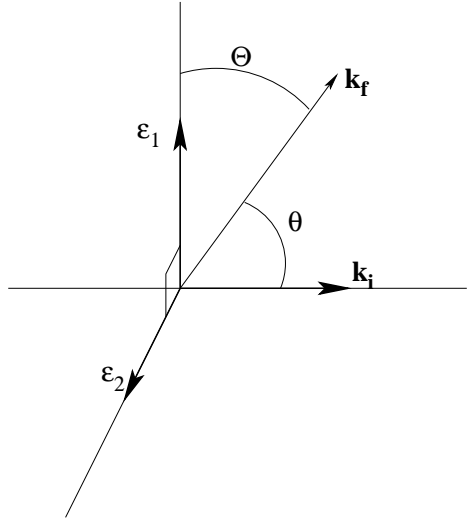}}
\end{center}
\end{figure}

Without loss of generality, we can carry out one more transformation
and define the initial photon direction to lie along the z-axis in the
electron frame. The x-y plane is decomposed into
$\bm{\varepsilon}_1$ and
$\bm{\varepsilon}_2$, where the initial
polarization is aligned with $\bm{\varepsilon}_1$. The scattered radiation
$\mathbf{k}_f$ makes an angle $\Theta$ with $\bm{\varepsilon}_1$
and $\theta$ with $\mathbf{k}_i$, as shown in Figure \ref{fig:scat_schem}.
For unpolarized incident light, we can define $\bm{\varepsilon}_1$ to
lie in the plane of $\mathbf{k}_i$ and $\mathbf{k}_f$, with
$\Theta+\theta=90^\circ$.

For photons
polarized along $\bm{\varepsilon}_1$, the angle-dependent
cross section $\sigma(\theta)$ is given by the dipole scattering
formula \citep{rybicki:04}:
\begin{equation}\label{eqn:cross_pol}
\left(\frac{d\sigma}{d\Omega}\right)_{\rm pol} = r_0^2\sin^2\Theta =
r_0^2 \cos^2\theta \cos^2\phi \, ,
\end{equation}
where $\phi$ is the standard azimuthal angle measured with respect to
$\bm{\varepsilon}_1$. 
Here the classical electron radius $r_0$ is given by
\begin{equation}
r_0 = \frac{e^2}{m_e c^2} = 2.82\times 10^{-13}\mbox{ cm}.
\end{equation}
For photons scattering in the $\mathbf{k}_i$-$\bm{\varepsilon}_2$
plane, the cross section is constant:
$d\sigma(\Theta=\pi/2)/d\Omega=r_0^2$. For unpolarized light, we
define $\bm{\varepsilon}_1$ as lying in the scattering plane, so the
scattering angle with respect to $\bm{\varepsilon}_2$ is
$\pi/2$. Because unpolarized
light is an equal combination of $\bm{\varepsilon}_1$- and
$\bm{\varepsilon}_2$-polarized photons, we can reproduce the
familiar cross section for unpolarized scattering: 
\begin{eqnarray}\label{eqn:cross_unpol}
\left(\frac{d\sigma}{d\Omega}\right)_{\rm unpol} &=& \frac{1}{2}
\left[\left(\frac{d\sigma(\Theta)}{d\Omega}\right)_{\rm pol} +
\left(\frac{d\sigma(\pi/2)}{d\Omega}\right)_{\rm pol}\right]
\nonumber\\
&=& \frac{1}{2}r_0^2(1+\cos^2\theta)\, .
\end{eqnarray}

For an arbitrary polarization degree $\delta$, the cross section can
be written as the sum of unpolarized light with weight $(1-\delta)$
and purely polarized light with weight $\delta$:
\begin{eqnarray}\label{eqn:cross_general}
\frac{d\sigma}{d\Omega} &=& \frac{1}{2}r_0^2(1-\delta)(1+\cos^2\theta) +
r_0^2 \delta(1-\sin^2\theta \cos^2\phi) \nonumber\\
&=& \frac{1}{2}r_0^2[(1+\cos^2\theta)-\delta\sin^2\theta\cos 2\phi]\, .
\end{eqnarray}
Given the angle-dependent cross section, we can either pick the
scattering angles $(\theta,\phi)$ directly from a distribution
function derived from (\ref{eqn:cross_general}), or alternatively,
pick the angles from a uniform distribution, and give the scattered
flux a weight based on the cross section. We compare these two methods
in Appendix \ref{section:app_kernel}.

\begin{figure}[t]
\caption{\label{fig:scat_schem2} Definitions of polarization axes in
  pre- and post-scattering coordinates. $\mathbf{k}_i$,
  $\mathbf{k}_f$, $\bm{\varepsilon}_\parallel$, and
  $\bm{\varepsilon}_\parallel'$ are all in the same plane, while
  $\bm{\varepsilon}_\perp$ and $\bm{\varepsilon}_\perp'$ are normal to
  that plane.}
\begin{center}
\scalebox{0.6}{\includegraphics*{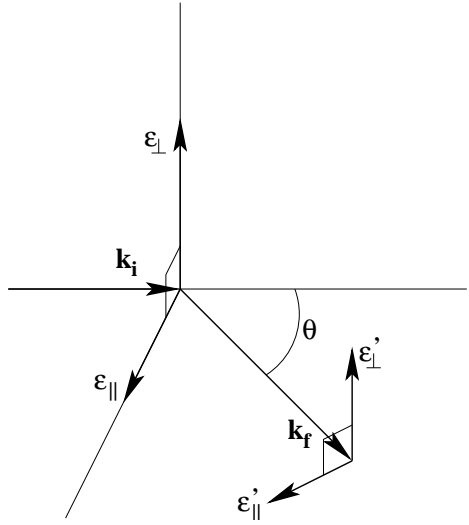}}
\end{center}
\end{figure}

Once the new photon direction is detemined, we need to calculate the
angle and degree of the post-scattered polarization. Here we follow
the Rayleigh matrix method described in \citet{chandra:60}. We define
yet another coordinate system with $\bm{\varepsilon}_3$ parallel to
$\mathbf{k}_i$, $\bm{\varepsilon}_\parallel$ in the scattering plane
defined by $\mathbf{k}_i$ and $\mathbf{k}_f$, and
$\bm{\varepsilon}_\perp$ normal to that plane. Likewise, we define a
post-scatter frame with $\bm{\varepsilon}_3'$ parallel to
$\mathbf{k}_f$, $\bm{\varepsilon}_\perp'=\bm{\varepsilon}_\perp$,
and $\bm{\varepsilon}_\parallel'$ in the scattering plane, but normal
to $\mathbf{k}_f$ (see Fig.\ \ref{fig:scat_schem2}). In this frame,
the initial polarization vector can be written $\mathbf{f}_i =
\cos\psi \bm{\varepsilon}_\parallel + \sin\psi
\bm{\varepsilon}_\perp$ and the final polarization is $\mathbf{f}_f =
\cos\psi' \bm{\varepsilon}_\parallel' + \sin\psi'
\bm{\varepsilon}_\perp'$. 

The standard Stokes parameters are given
by the intensity $I$, $Q=\delta I\cos 2\psi$, $U=\delta I\sin 2\psi$,
and $V=0$ 
(electron scattering never leads to circularly polarized
light). We further define
\begin{subequations}\label{eqn:I_vector}
\begin{eqnarray}
I_\parallel &\equiv& \frac{1}{2}(I+Q) = \frac{1}{2}(1-\delta)I +
\delta I \cos^2\psi \\
I_\perp &\equiv& \frac{1}{2}(I-Q) = \frac{1}{2}(1-\delta)I +
\delta I \sin^2\psi \\
\mathbf{I} &\equiv& [I_\parallel, I_\perp, U, V]
\end{eqnarray}
\end{subequations}
and the Rayleigh scattering phase matrix 
\begin{equation}
\mathbf{R} = \begin{pmatrix}
\cos^2\theta & 0 & 0 & 0 \\
0 & 1 & 0 & 0 \\
0 & 0 & \cos\theta & 0 \\
0 & 0 & 0 & \cos\theta 
\end{pmatrix}\, .
\end{equation}
Then the scattered Stokes parameters are given simply by
$\mathbf{I}'=\mathbf{RI}$, $I' = I_\parallel'+I_\perp'$, and $Q' =
I_\parallel'-I_\perp'$. Note that the cross section
(\ref{eqn:cross_general}) can be reproduced by writing
\begin{equation}
I' = \cos^2\theta I_\parallel+I_\perp =  
\frac{1}{2}(1-\delta)I(1+\cos^2\theta)+\delta I (\cos^2\theta \cos^2\psi+\sin^2\psi)\, ,
\end{equation}
giving
\begin{equation}\label{eqn:I_Iprime}
\frac{I'}{I} = \frac{1}{2}(1-\delta)(1+\cos^2\theta)+
\delta (1-\sin^2\theta \cos^2\psi)\, ,
\end{equation}
now with $\psi$ taking the place of $\phi$ from equation
(\ref{eqn:cross_general}). 

Lastly, $\mathbf{f}_f$ is constructed by
\begin{subequations}
\begin{eqnarray}
\delta' &=& \frac{\sqrt{Q'^2+U'^2}}{I'} \\
\psi' &=& \frac{1}{2}\tan^{-1}(U',Q') \\
\mathbf{f}_f &=& \cos\psi' \bm{\varepsilon}_\parallel' + \sin\psi'
\bm{\varepsilon}_\perp' \, .
\end{eqnarray}
\end{subequations}
At this point, the polarization vector and photon 4-momentum are
transformed back into corona fluid frame, then to the coordinate
frame, and then the geodesic
propagation continues as before, until the photon packet scatters
again, is absorbed by the black hole, or reaches a distant observer. 

During this scattering process, the photon packet's array of
frequencies had to be adjusted three times: once when transforming
from the fluid frame to the electron rest frame, once when losing
energy to the electron recoil, and once when transforming back to the
fluid frame. The first and last transformations are simple Lorentz
boosts, and the frequency scales like the photon energy: $\nu'/\nu =
p^{t'}/p^t$, with $p^{t'}=\Lambda_{\,\, \mu}^{t'} p^\mu$. For the scattering
losses, we need to
scale the frequency bins such that the number of photons in each bin
is conserved, while losing energy according to the Compton recoil
formula: 
\begin{equation}
E_f = \frac{E_i}{1+\frac{E_i}{m_ec^2}(1-\cos\theta)}\, .
\end{equation}
Thus the frequency scales like
\begin{equation}
\frac{\nu'}{\nu} =
\left[1+\frac{h\nu}{m_ec^2}(1-\cos\theta)\right]^{-1} 
\end{equation}
and the size of each bin scales like
\begin{equation}
\frac{d\nu'}{d\nu} =
\left[1+\frac{h\nu}{m_ec^2}(1-\cos\theta)\right]^{-2} \, .
\end{equation}
The number of photons in each bin $dN_\nu = F_\nu d\nu/(h\nu)$ is
conserved in the scattering event, so we find that the effect of
Compton recoil on the spectral luminosity is
\begin{equation}
\frac{F_\nu'}{F_\nu} =
\left[1+\frac{h\nu}{m_ec^2}(1-\cos\theta)\right] \, .
\end{equation}

At very high energies $h\nu \gg m_ec^2$, this leads to a ``pile up''
of photons and large peaks in the photon packet spectrum. In reality,
this effect would be mitigated by incorporating Kline-Nishina cross
sections, which decrease with energy, yet are incompatible with our
photon packet approach that treats all photons as identical regardless
of frequency. In practice, we are generally interested in problems
where the characteristic photon energies are significantly below
$m_ec^2$, so the photon pile up is rarely an issue. 

While some energy is lost to Compton recoil in the electron frame, the
more typical effect is {\it inverse-Compton} scattering, where energy is transferred
from the electrons to the photons. For electrons with Lorentz factors
$\gamma$ in the fluid frame and low-energy photons with $h\nu/m_ec^2
\ll \gamma^2-1$, the ratio of energies of the photons before
scattering, in the rest frame of the electron, and after scattering
is roughly $1:\gamma:\gamma^2$ \citep{rybicki:04}. For coronal
electrons with temperature $\sim 140$ keV, low-energy seeds will, on
average, double their energy after every scattering event, making
inverse-Compton a very efficient radiative process.

\subsection{Disk Scattering}\label{section:disk_scattering}
At each step along the geodesic trajectory, we determine whether or
not the photon packet has crossed the photosphere surfaces
$\Theta_{\rm top}(r,\phi)$ or $\Theta_{\rm bot}(r,\phi)$. If it has
crossed this boundary, we follow a procedure similar to that described
above for coronal scattering. First, we use the conserved quantities
$\kappa_{\rm wp}$, $\mathbf{f}\cdot\mathbf{f}=1$, and
$\mathbf{f}\cdot\mathbf{k}=0$ to solve for the polarization vector
$\mathbf{f}$ in the coordinate frame. Then we transform $\mathbf{f}$
and $\mathbf{k}$ into the local fluid frame of the photosphere tetrad
$\mathbf{e}_{(\bar{\mu})}$, with $\mathbf{e}_{(\bar{z})}$ normal to
the disk surface, as in equation (\ref{eqn:surface_tetrads}). 

In this frame, the scattering off the disk surface is calculated using
the analytic expressions for reflection off a diffuse semi-infinite
plane, derived by Chandrasekhar and given in table XXV of
\citet{chandra:60}. As in equation (\ref{eqn:I_vector}) above, we can write
the incoming photon beam as a vector of Stokes parameters for the flux
$\mathbf{I}=[I_\parallel, I_\perp, U]$ ($V=0$ for
linearly-polarized light, the only relevant case for our
scattering-dominated systems). Then the outgoing intensity is given by 
\begin{equation}\label{eqn:diffuse_reflection}
\mathbf{I}'(\mu,\varphi) = \begin{pmatrix} I'_\parallel \\ I'_\perp \\ U'
\end{pmatrix} = \frac{1}{4\mu\mu_0} \mathbf{Q} \mathbf{S}(\mu,\varphi;
\mu_0, \varphi_0) \begin{pmatrix} I_\parallel \\ I_\perp \\ I_U
\end{pmatrix}\, ,
\end{equation}
where ($\mu_0$, $\varphi_0$) are the incident angles in the fluid
frame with $\mu_0=|k_{\bar{z}}|$, ($\mu$, $\varphi$) are the outgoing
angles, and $\mathbf{Q}$ and $\mathbf{S}$ are the transfer matrices
defined in Section 70.3 of \citet{chandra:60}. Unlike the coronal
scattering case, where we use the differential cross section (\ref{eqn:cross_general})
to determine the post-scatter angles, for diffuse reflection off the
disk, we simply choose a random angle $(\mu, \varphi)$ from a uniform
distribution and then weight the outgoing intensity by $I'/I$ from
equation (\ref{eqn:diffuse_reflection}). Thus any individual
reflection does not conserve photon number, but the angle-averaged
process does. From $\mathbf{I}'$, we are able to reproduce $\delta'$, $\psi'$, and
thus $\mathbf{f}'$ and $\mathbf{k}'$ as above, which are then
transformed back into the coordinate frame and continue their geodesic
propagation through the corona. 

This method for diffuse reflection can be checked against coronal 
scattering experiments where we scatter incoming photons off a
semi-infinite plane of free electrons. We find excellent agreement
between the analytic and numerical approaches, as
shown below in Section \ref{section:tests}.

As with the coronal scattering, high energy photons can lose energy to
Compton recoil off the electrons in the cool disk, leading to the
reflection hump seen in many AGN observations. While this process
is technically angle-dependent, as a simplification, we average over
all incoming and outgoing angles, as well as the number of individual
scatterings typically responsible for diffuse reflection $(N_{\rm
  scat}\approx 3)$, and use the recoil formula
\begin{equation}\label{eqn:recoil}
\frac{\nu'}{\nu} =
\left(1+3\frac{h\nu}{m_ec^2}\right)^{-1} \, .
\end{equation}
This energy lost by the photons can then be reprocessed by the disk
and emitted at thermal energies.

\section{NUMERICAL TESTS}\label{section:tests}

Here we present a number of test problems to verify \pand's accuracy
and reliability. We begin with vacuum transport of geodesics from the
disk to a distant observer. To test the tetrad construction methods
outlined in Section \ref{section:phototetrad}, we calculate the
relativistic broadening of iron lines from a thin disk around a Kerr
black hole, comparing the emitter-to-observer and observer-to-emitter
paradigms. The observer-to-emitter approach is well-known in the
literature \citep{rauch:94,broderick:03,broderick:04}. It is also
relatively straight-forward conceptually, since it 
doesn't require the use of any tetrads or proper area
calculations. One simply shoots rays backwards from a distant
observer, and integrate the geodesic path until the ray crosses the
midplane, where the fluid 4-velocity can be determined analytically as
in \citet{novikov:73}. This gives the redshift of the emission line as
seen by the observer, and the spectrum is given by the invariant
$I_\nu/\nu^3$. 

\begin{figure}[t]
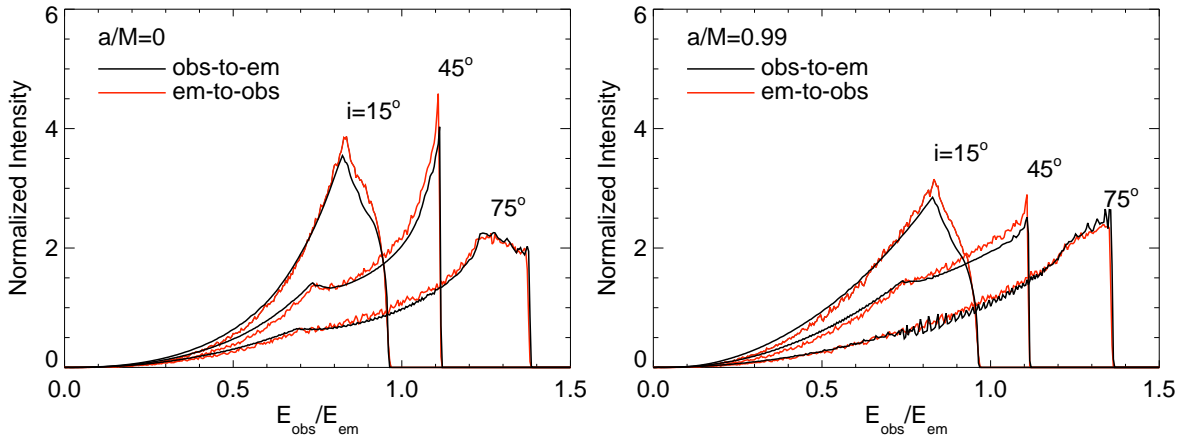

\caption{\label{fig:ironline_test} Comparison of observer-to-emitter
  and emitter-to-observer ray-tracing paradigms for a relativistically
  broadened emission line. The disk extends from $R_{\rm out}=15M$ all
  the way to the horizon, and the emissivity profile scales like
  $r^{-3}$.}
\begin{center}
\scalebox{0.5}{\includegraphics*{ironline_test_a0.epsi}}
\scalebox{0.5}{\includegraphics*{ironline_test_a99.epsi}}
\end{center}
\end{figure}

In Figure \ref{fig:ironline_test}, we show the shape of a
relativistically broadened emission line as viewed by observers at
different inclination angles for the spin values $a/M=0$ and $0.99$. In all
cases, the emissivity profile scales like $I\sim r^{-3}$ and the outer
disk is truncated at $r=15M$. The disk extends all the
way into the horizon, with the fluid velocity inside the ISCO
determined by conserving the energy and angular momentum at the ISCO,
and solving for $u^r$ from the relation $u_\mu u^\mu=-1$. For the
observer-to-emitter calculation, we use the same ray-tracing code
described in \citet{schnittman:04}, with $10^7$ photons evenly spaced
in the image plane for each inclination. We find
excellent agreement in all cases, validating our emitter-to-observer
techniques, at least for planar test-particle orbits.

\begin{figure}[t]
\caption{\label{fig:ironline_converge} Error estimate $\varepsilon$ as
  a function of photon number for a relativistically
  broadened iron line, defined in equation (\ref{eqn:error}). 40
  inclination bins were used.}
\begin{center}
\scalebox{0.5}{\includegraphics*{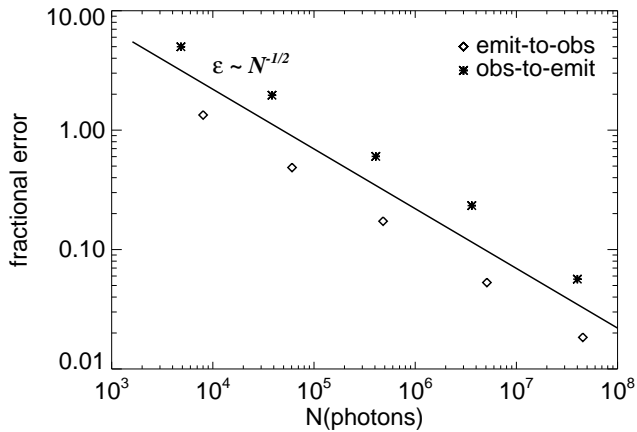}}
\end{center}
\end{figure}

This test in turn naturally leads to a simple convergence test for our
Monte Carlo code. Integrating over energy and observer inclination angle
$i$, we can apply the following metric to estimate the error due
to the use of a finite number of photons:
\begin{equation}\label{eqn:error}
\varepsilon = \frac{\left[\int d\cos i \int dE (I_{\rm lo}(E)-I_{\rm
      hi}(E))^2 \right]^{1/2}}{\left[\int d\cos i \int dE I_{\rm
      hi}^2(E) \right]^{1/2}}\, ,
\end{equation}
with $I_{\rm lo}(E)$ the spectrum calculated at low resolution,
compared to the theoretically perfect spectrum $I_{\rm hi}(E)$
calculated at high resolution. As expected for a Monte Carlo
calculation, we find that the total error scales with photon number
like $N^{-1/2}$, as shown in Figure \ref{fig:ironline_converge}. This
is consistent with similar spectral calculations done with the Monte
Carlo radiation code {\tt grmonty} \citep{dolence:09}. Also shown in
Figure \ref{fig:ironline_converge} are the errors $\varepsilon(N)$ for
the observer-to-emitter approach, using a total of 40 inclinations for
both cases. We note that the emitter-to-observer method is more than a
factor of two more efficient for the same calculation. This is because
we can selectively shoot more photons from the inner regions of the
disk, but in the reverse method, the photons are distributed evenly in
the image plane (this uniform distribution is not strictly necessary;
e.g., \citet{noble:07} use an adaptive grid to improve resolution in {\tt
bothros}). 

The next test is similar, but also includes polarization
effects. Instead of an emission line with $I(r)\sim r^{-3}$, we use
the diluted thermal spectrum for a Novikov-Thorne (NT) disk with an inner edge at the
ISCO. The emission has the polarization and limb darkening appropriate
for a scattering-dominated atmosphere \citep{chandra:60}. For the
observer-to-emitter approach, in addition to utilizing the
$I_\nu/\nu^3$ invariance, we also parallel-transport two polarization
basis vectors corresponding to the two axes in the observer plane
normal to the photon propagation direction. Then, when the ray
intersects with the disk, we calculate a local tetrad in
order to determine the local angle of incidence and thus degree of
polarization. The direction of polarization is projected onto the
parallel-transported basis vectors to give the observed angle at
infinity. 

\begin{figure}[t]
\caption{\label{fig:polar_images} Comparison of observer-to-emitter
  and emitter-to-observer polarized images for a NT disk with
  polarization given by a scattering-dominated atmosphere. 
  The disk extends from an inner edge at the ISCO out to $R_{\rm
  out}=15M$. The black hole has spin $a/M=0.99$ and the observer is at
  an inclination of $i=75^\circ$. The intensity color scale is
  logarithmic and the polarization vectors are linearly proportional
  to the local degree of polarization, as observed at infinity.}
\begin{center}
\scalebox{0.6}{\includegraphics*[75,450][410,675]{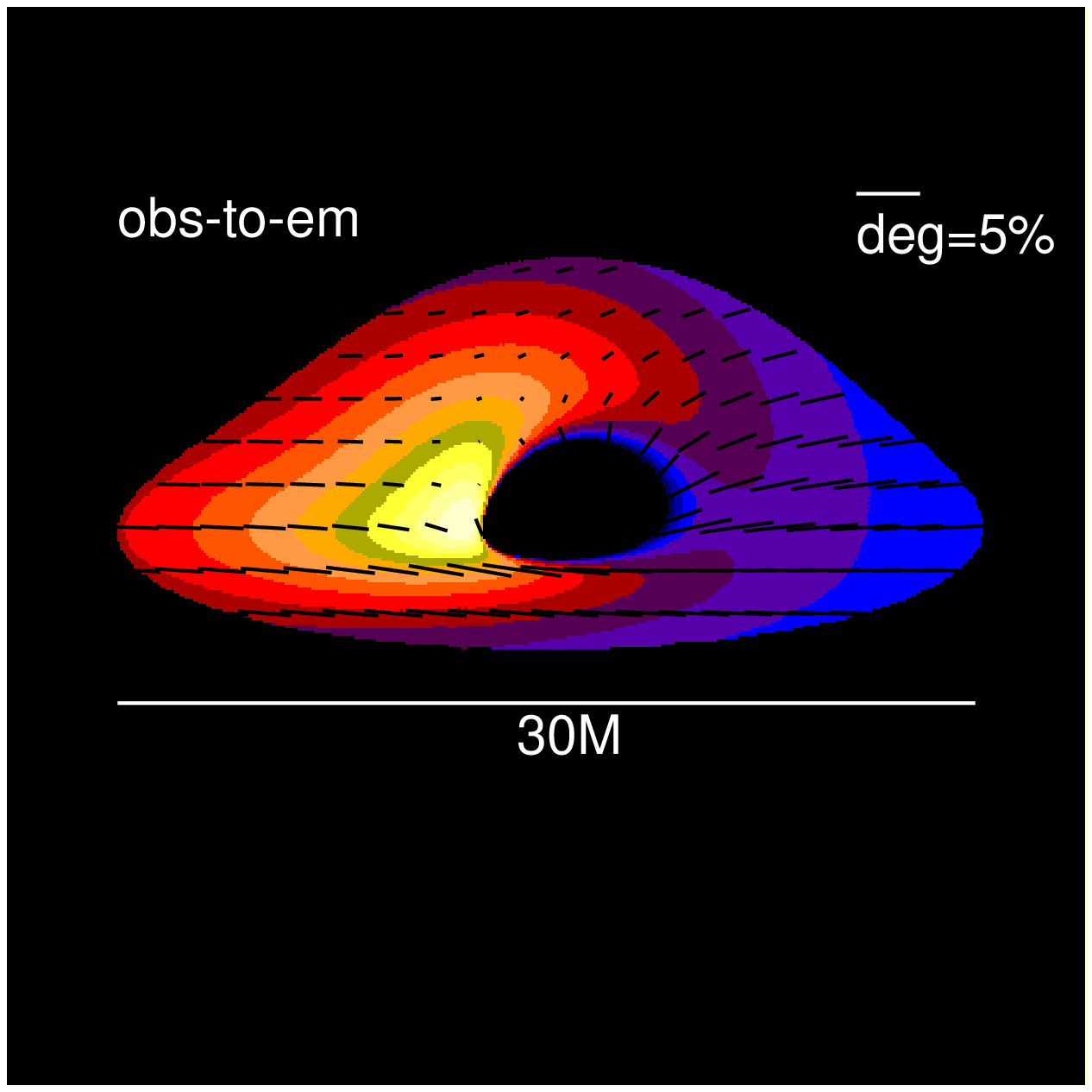}}
\scalebox{0.6}{\includegraphics*[75,450][410,675]{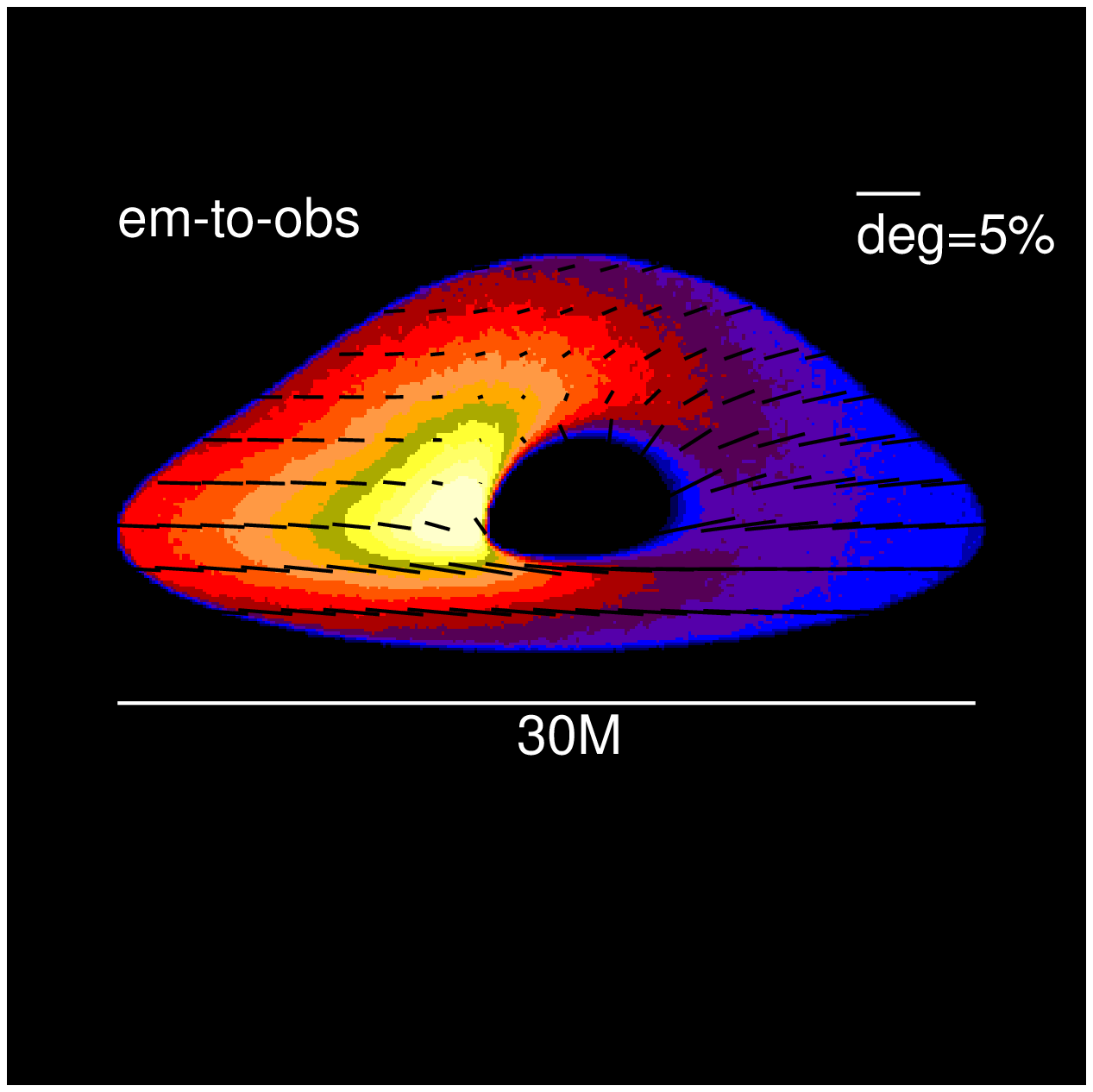}}
\end{center}
\end{figure}

The two approaches give identical results, as shown in the images in
Figure \ref{fig:polar_images}, for a Kerr black hole with spin
$a/M=0.99$, $R_{\rm out}=15M$, and observer inclination angle
75$^\circ$. The color code is logarithmic in total intensity and
covers four orders of magnitude, and the
small vectors scale linearly with degree of polarization. For the
purposes of comparison, we have not included returning
radiation here, despite the important effect it has on the
polarization signal \citep{agol:00,schnittman:09}. In fact, it is
precisely due to the critical importance of returning radiation that
we were forced to employ the emitter-to-observer approach in
\citet{schnittman:09,schnittman:10}. 

\begin{figure}[t]
\caption{\label{fig:degree_angle} Polarization degree and angle as a
  function of photon energy for a black hole with spin $a/M=0.99$,
  luminosity $0.1L_{\rm Edd}$, and mass $10M_\odot$. The disk extends
  from the ISCO out to $r=15M$. The
  observer-to-emitter (solid curves) and emitter-to-observer
  (diamonds) methods agree closely over a range of inclinations.}
\begin{center}
\scalebox{0.45}{\includegraphics*{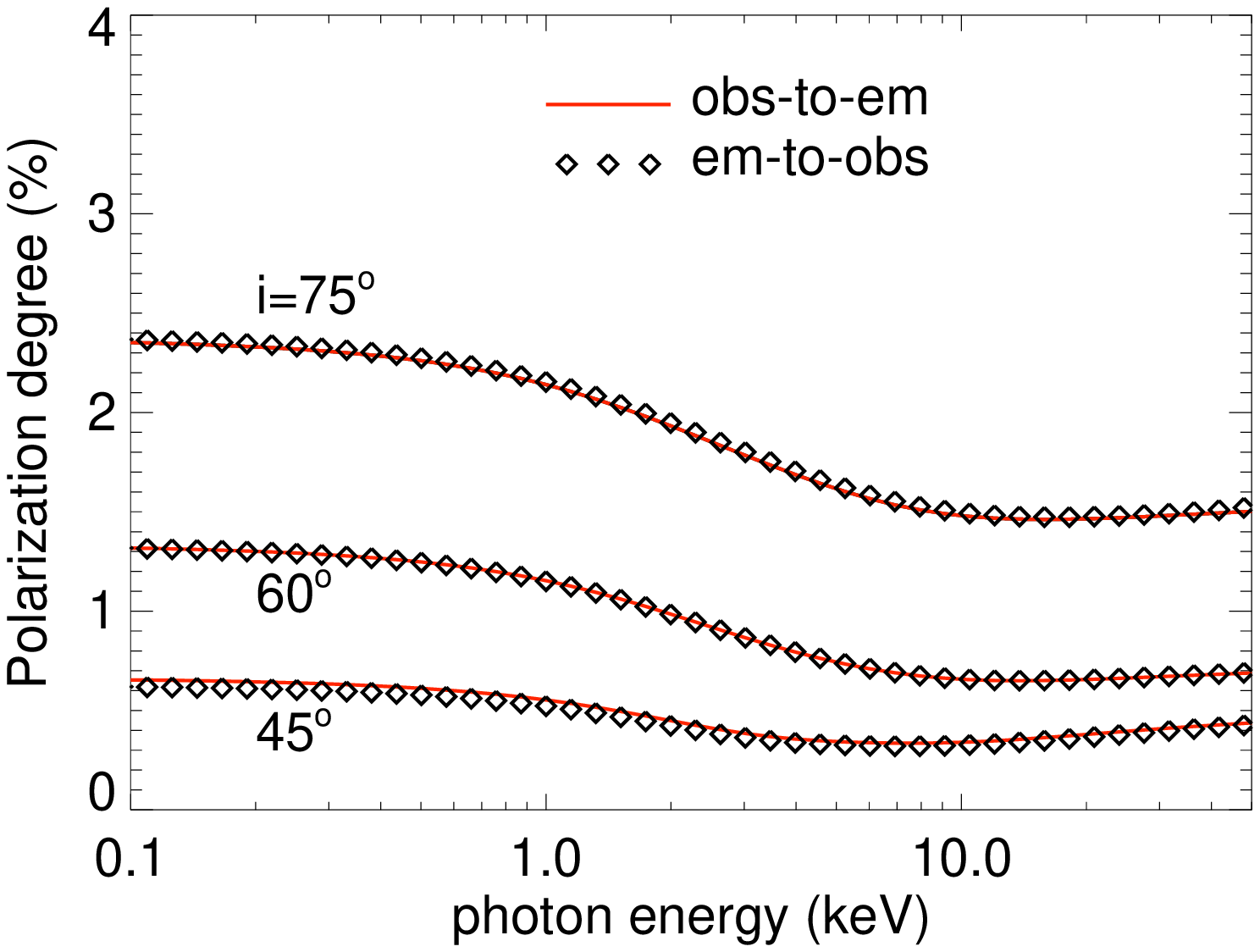}}
\scalebox{0.45}{\includegraphics*{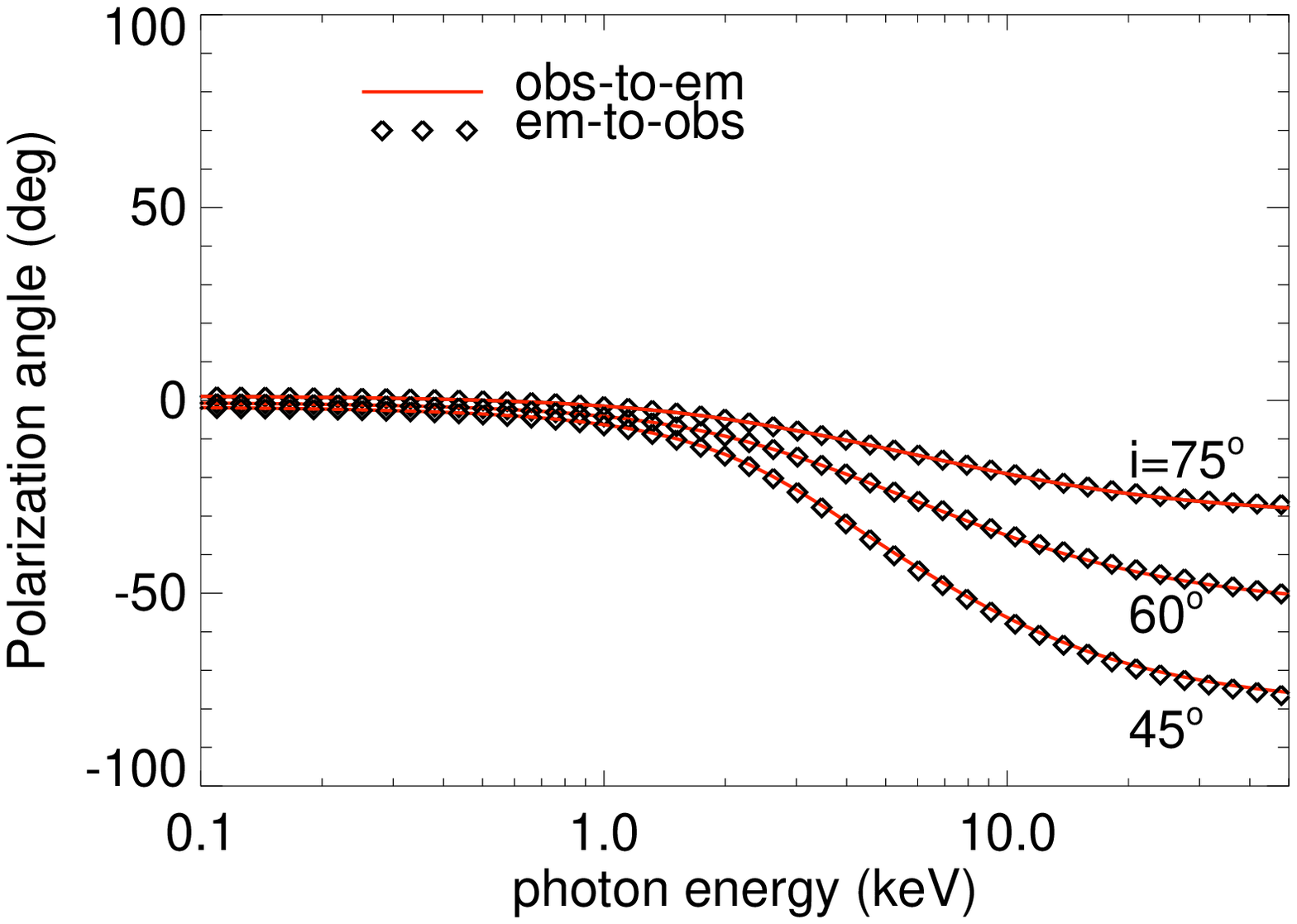}}
\end{center}
\end{figure}
In Figure
\ref{fig:degree_angle} we show the observables of polarization degree
and angle as a function of energy for a range of inclination angles,
assuming an Eddington-scaled accretion rate of $\dot{m}=0.1$ and black
hole mass $M=10M_\odot$. Again, we find excellent agreement between
the emitter-to-observer and observer-to-emitter methods. 

\begin{figure}[t]
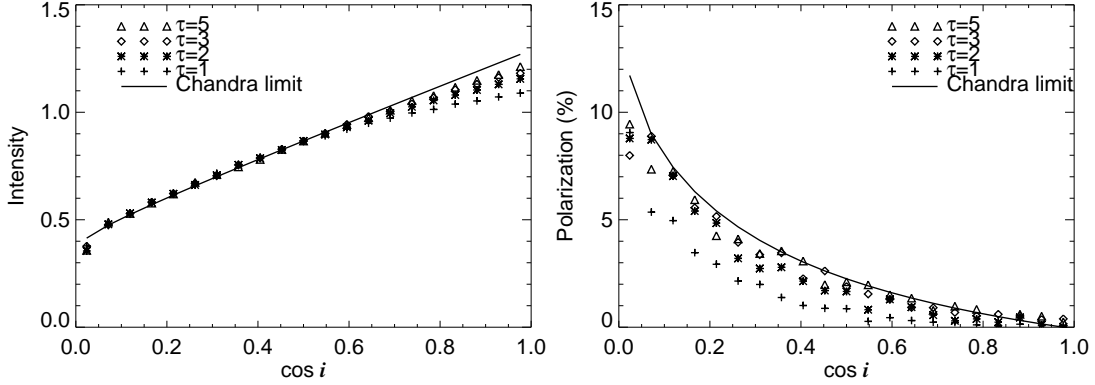

\caption{\label{fig:chandra1} Outgoing intensity and degree of
  polarization for radiation emitted from a scattering-dominated
  atmosphere, as a function of inclination, and for a range of
  corona optical depths. The face-on orientation is
  unpolarized due to symmetry.}
\begin{center}
\scalebox{0.45}{\includegraphics*{chandra_intensity.epsi}}
\scalebox{0.45}{\includegraphics*{chandra_degree.epsi}}
\end{center}
\end{figure}
Next, we move on to testing the coronal scattering algorithms. We
focus on a plane-parallel geometry with an optically thick disk
covered by a corona with variable optical depth $\tau$ and electron
temperature $T_e$. In Figure \ref{fig:chandra1} we show the effects of
a scattering atmosphere on the emergent flux and polarization as a
function of angle. The seed photons are emitted isotropically from the
disk surface with zero polarization, then scatter through a cold
corona. Photons that scatter back to the disk are reflected via the
diffuse scattering formula of equation
(\ref{eqn:diffuse_reflection}). In the limit of $\tau\to\infty$, we
reproduce the limb-darkening and horizontal polarization results from
\citet{chandra:60}, Table XXIV.

\begin{figure}[t]
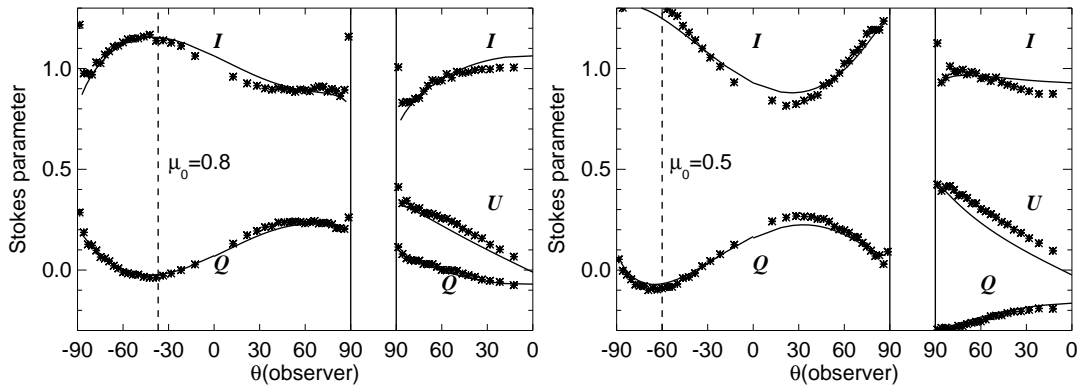

\caption{\label{fig:chandra2} Stokes parameters for scattering of
  unpolarized light off an optically thick atmosphere. See text for
  description. Compare with Figures 24 and 25 of \citet{chandra:60}.}
\begin{center}
\scalebox{0.45}{\includegraphics*{chandra_reflection08.epsi}}
\scalebox{0.45}{\includegraphics*{chandra_reflection05.epsi}}
\end{center}
\end{figure}
In Figure \ref{fig:chandra2} we carry out a similar scattering
experiment, but with the seed photons incident from above the disk
along a single direction. Setting $\tau=10$, we should reproduce the
analytic diffuse reflection expressions of \citet{chandra:60}, shown
there in Figures 24--25 for an incident unpolarized beam with
$\cos\theta_0=0.8, 0.5$ and $\varphi_0=0$. Following
\citet{chandra:60}, we plot the Stokes parameters $I$, $Q$, and $U$ as
a function of reflection angle, normalized to the incident
intensity. The asterisks are the Monte Carlo calculation, and the solid
curves are the analytic predictions.

On the left-hand side of each plot, we show the
polarization as a function of $\theta$ for
$\varphi-\varphi_0=0^\circ,\pm 180^\circ$. The value of $\theta_0$ is
designated with a vertical dashed line. Negative values of $\theta$
correspond to photons reflected back in the general direction of the
incident photons, i.e., $\varphi-\varphi_0=0$. Thus we see a natural
peak in the intensity
corresponding to backscattering as in equation
(\ref{eqn:cross_unpol}). Similarly, the degree of polarization is
maximized for $90^\circ$ scattering, and oriented in the plane of the
disk $(Q>0)$. 

On the right-hand side of each plot, we show the Stokes parameters for
photons scattered with $\varphi-\varphi_0=\pm 90^\circ$. In this case,
the planar symmetry is broken and we find a non-zero value of
$U$. Again, the degree of polarization is maximized for scattering
angles near $90^\circ$. 

Lastly, we test the inverse-Compton effects of a hot corona by
reproducing the AGN-type spectra of \citet{poutanen:96}. The seed
photons are again isotropic and unpolarized, with a blackbody
spectrum with $T_{\rm bb}=10$ eV. When reflecting off the cold disk,
we implement the Compton recoil losses of equation
(\ref{eqn:recoil}). Following \citet{poutanen:96}, we also include
atomic absorption in the disk with an extremely simple toy model based
on the photoelectric cross sections of \citet{morrison:83}.

\begin{figure}[t]
\caption{\label{fig:poutanen1} Spectra and polarization of flux from a
  disk and corona with a slab geometry corresponding to an AGN
  accretion disk, as described in the text. The solid
  curves correspond to the total flux, while the (dotted, dashed,
  dot-dashed, triple-dot-dashed, and long-dashed) curves correspond to
  $N_{\rm scat} =  (0, 1, 2, 3, \ge 5)$. For clarity, only the solid
  curves are shown for the polarization degree. Compare with Figure 5 of
  \citet{poutanen:96}.}
\begin{center}
\scalebox{0.48}{\includegraphics*[55,415][540,700]{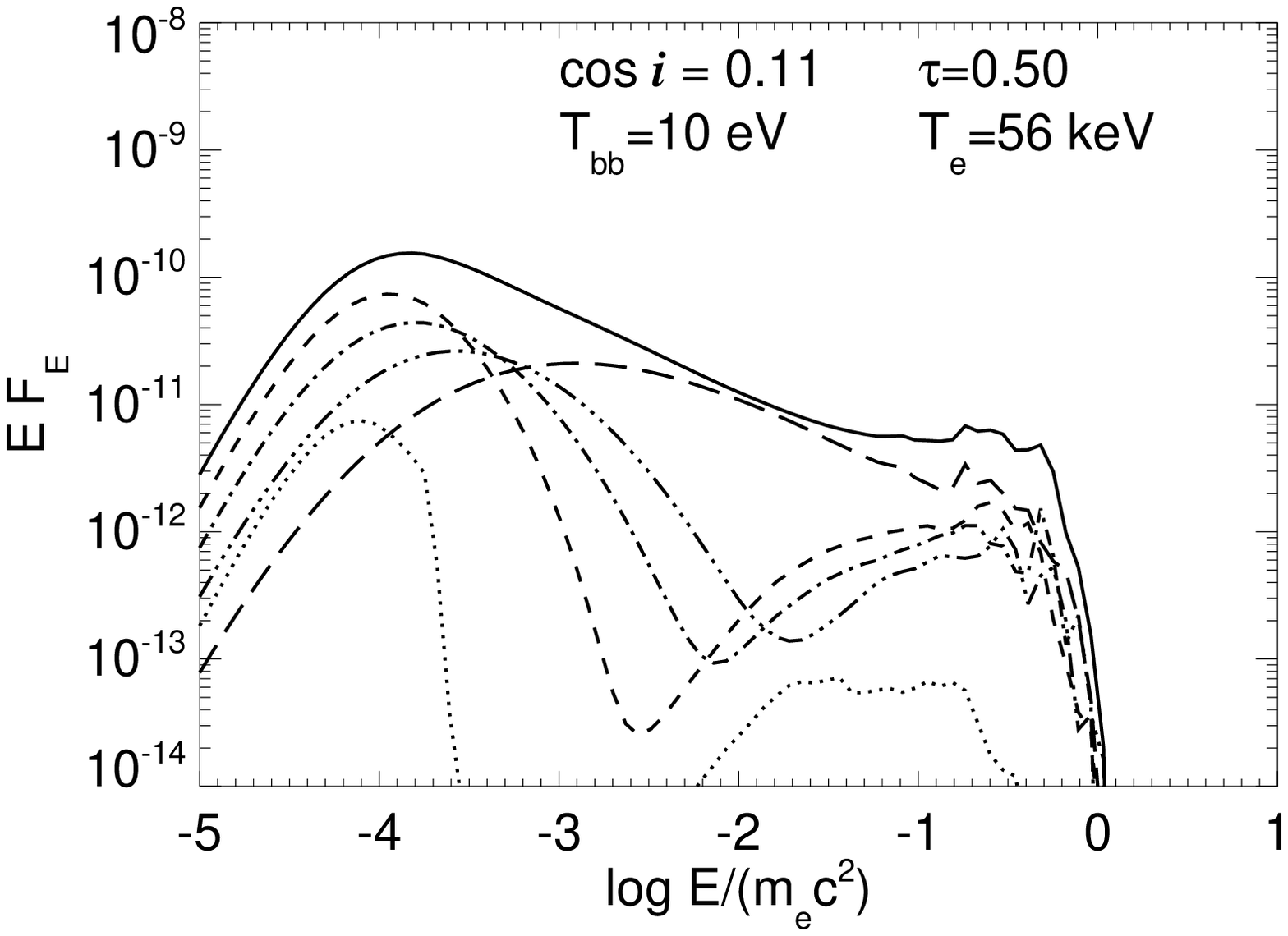}}
\scalebox{0.48}{\includegraphics*[145,415][540,700]{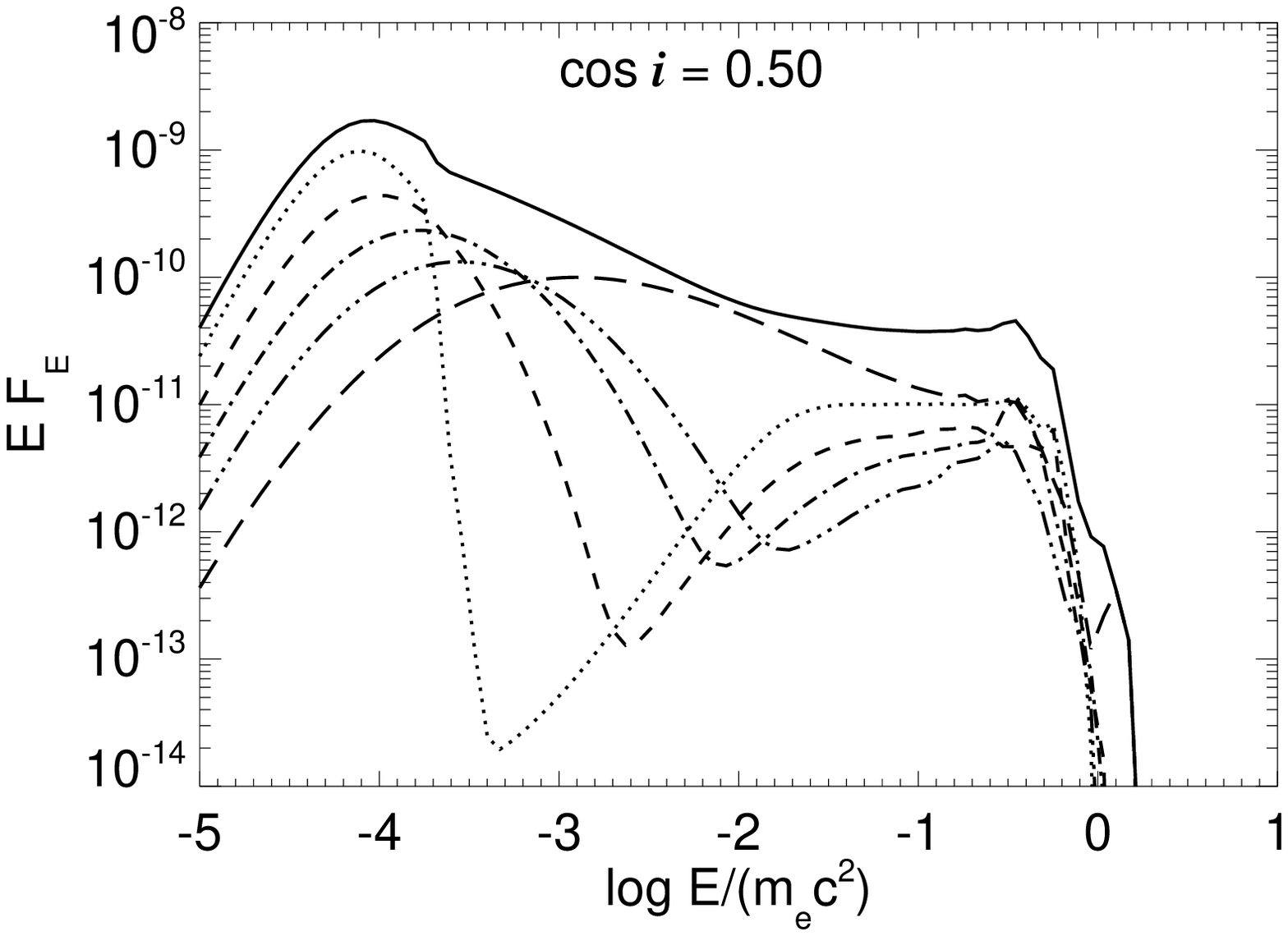}}
\scalebox{0.48}{\includegraphics*[55,360][540,700]{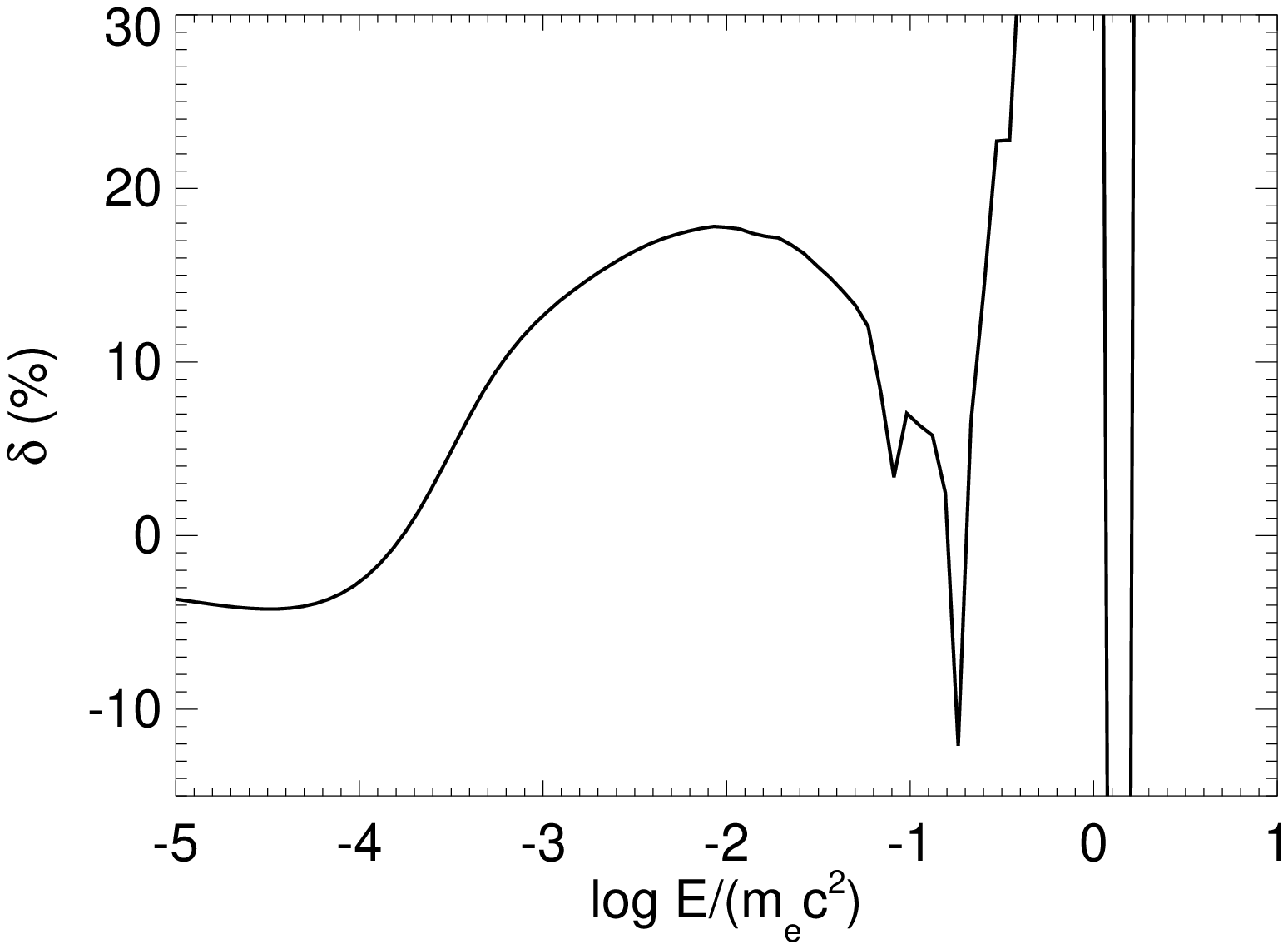}}
\scalebox{0.48}{\includegraphics*[145,360][540,700]{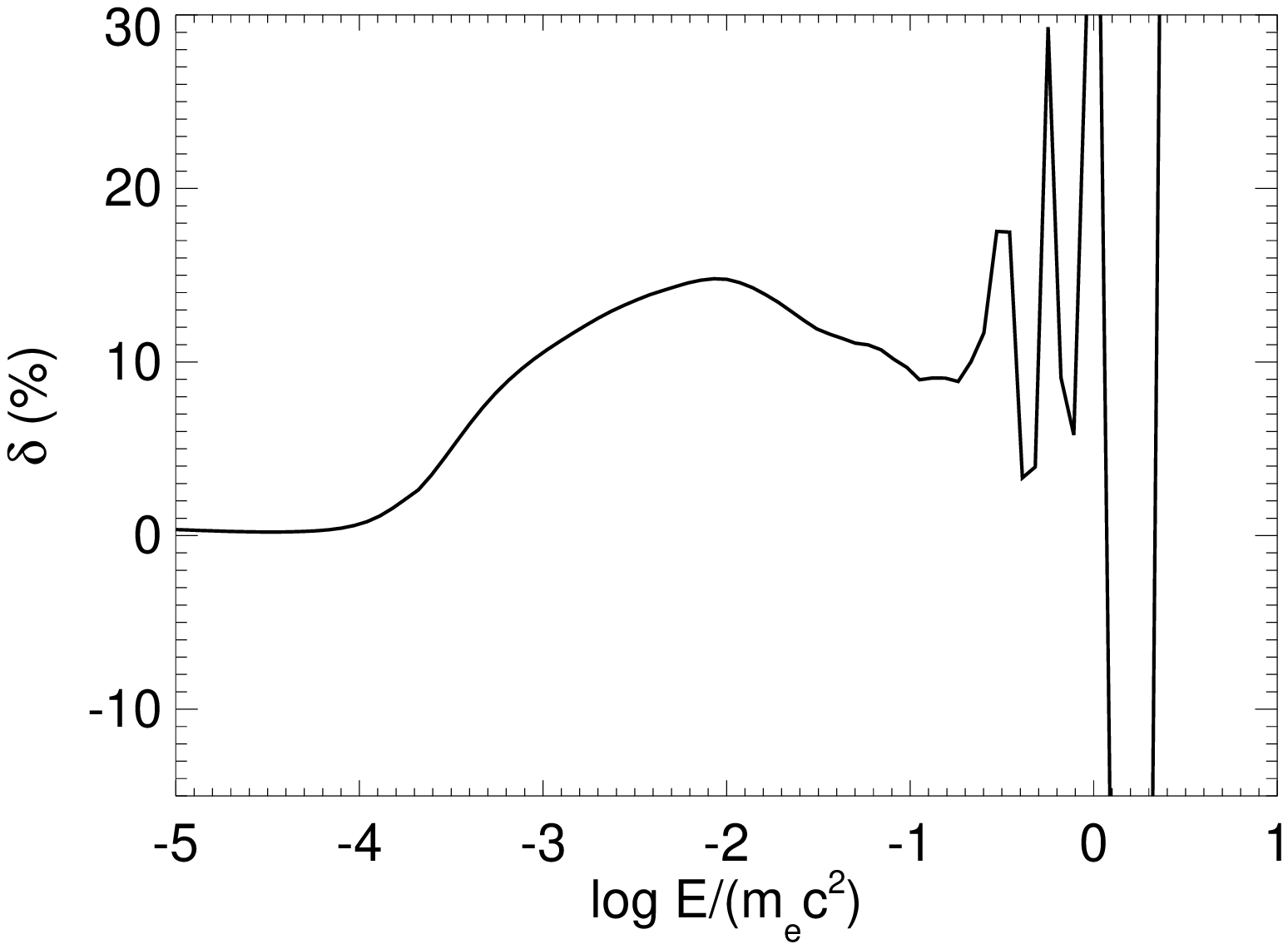}}
\end{center}
\end{figure}

\begin{figure}[t]
\caption{\label{fig:poutanen2} Same as Figure \ref{fig:poutanen1}, but
  for $T_e=352$ keV and $\tau=0.05$. Compare with Figure 6 of
  \citet{poutanen:96}.}
\begin{center}
\scalebox{0.48}{\includegraphics*[55,415][540,700]{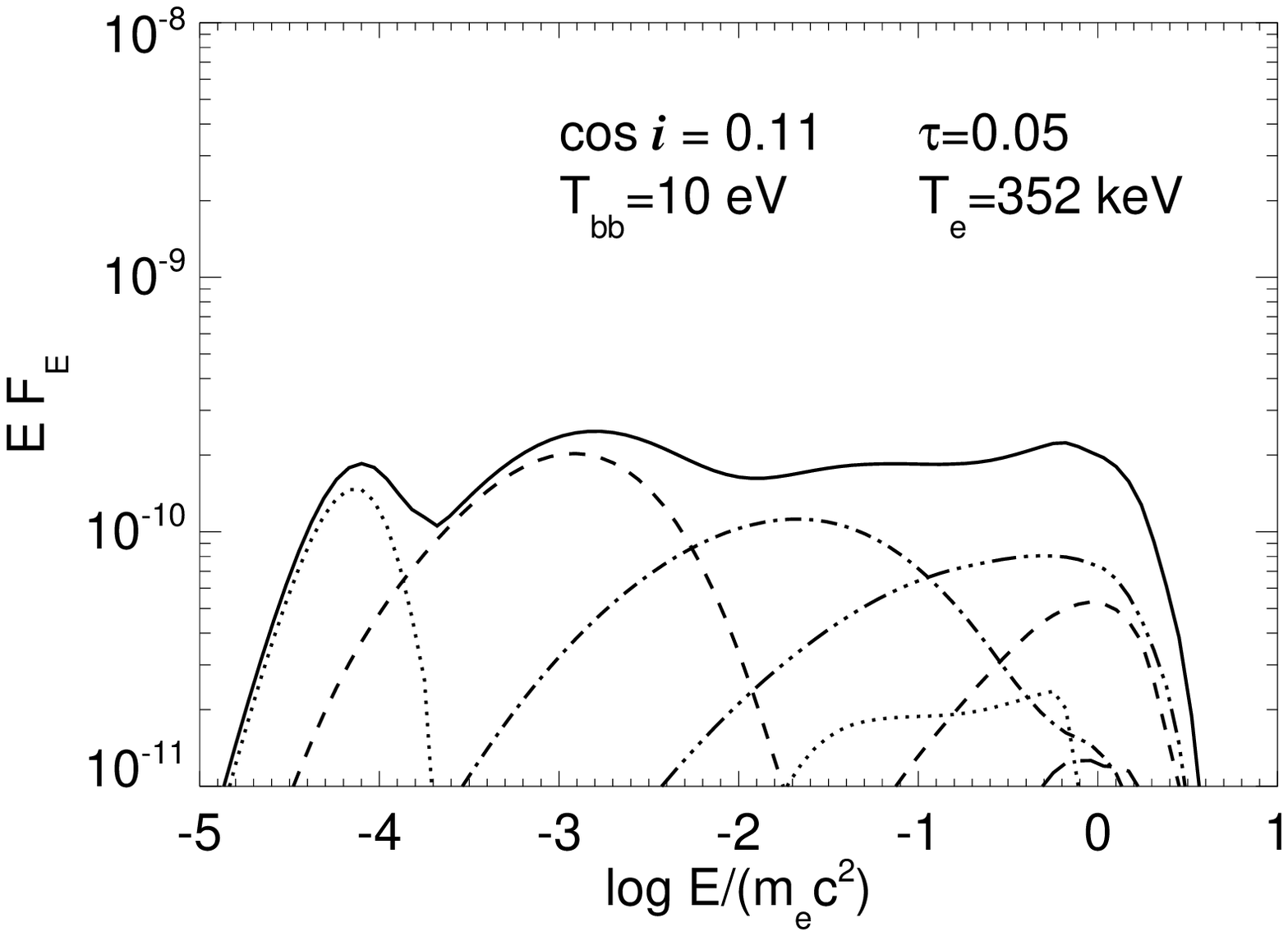}}
\scalebox{0.48}{\includegraphics*[145,415][540,700]{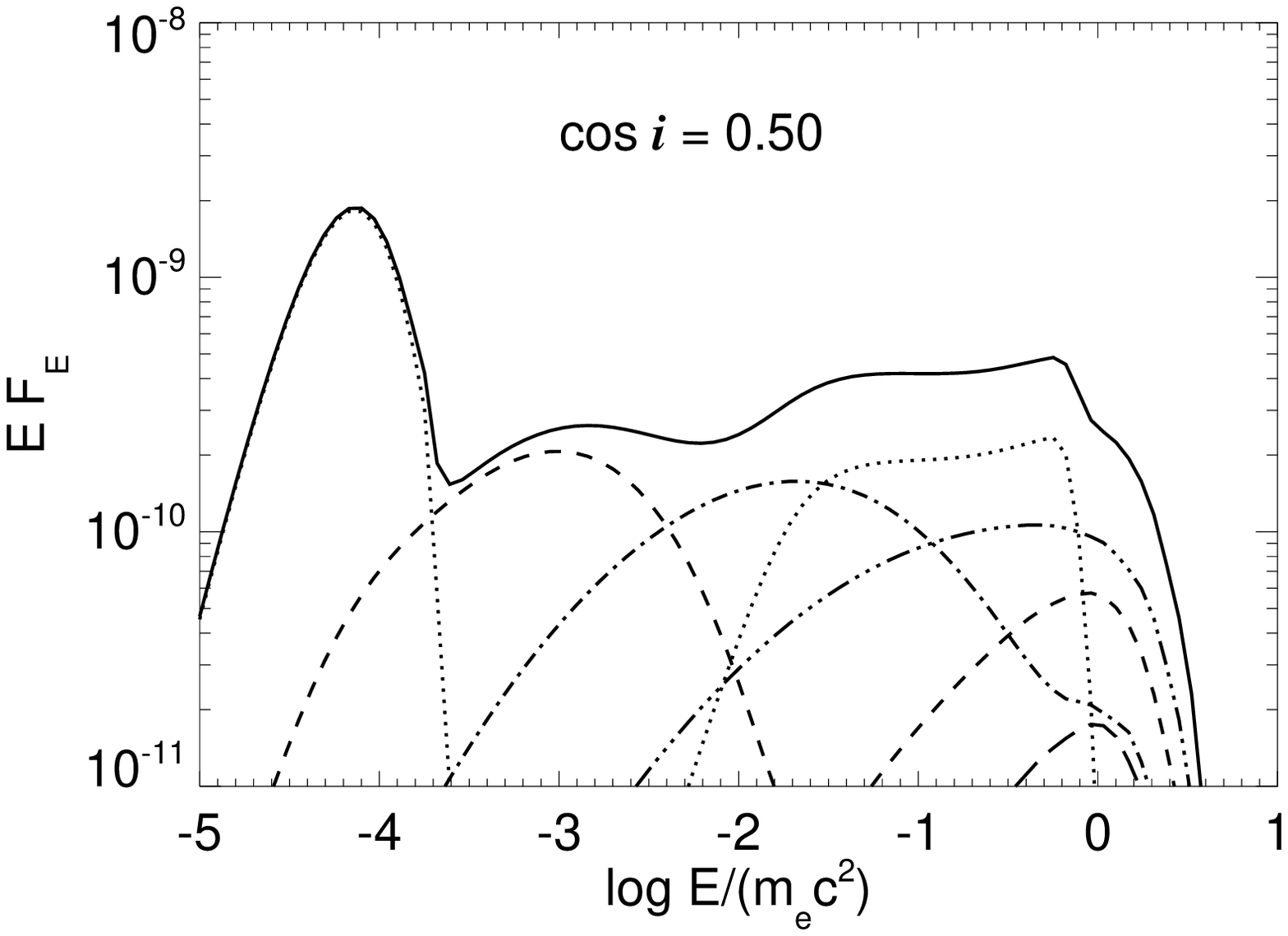}}
\scalebox{0.48}{\includegraphics*[55,360][540,700]{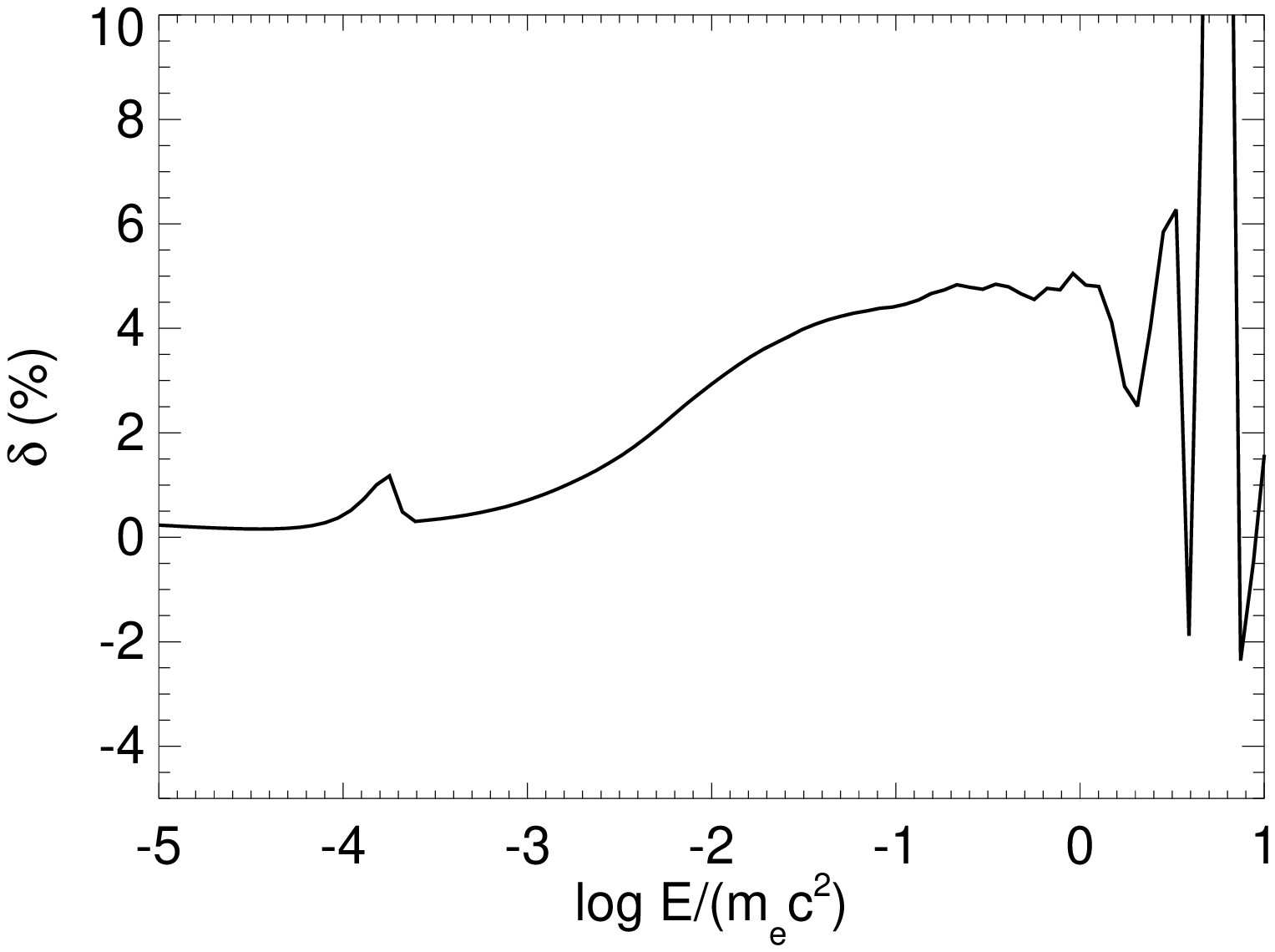}}
\scalebox{0.48}{\includegraphics*[145,360][540,700]{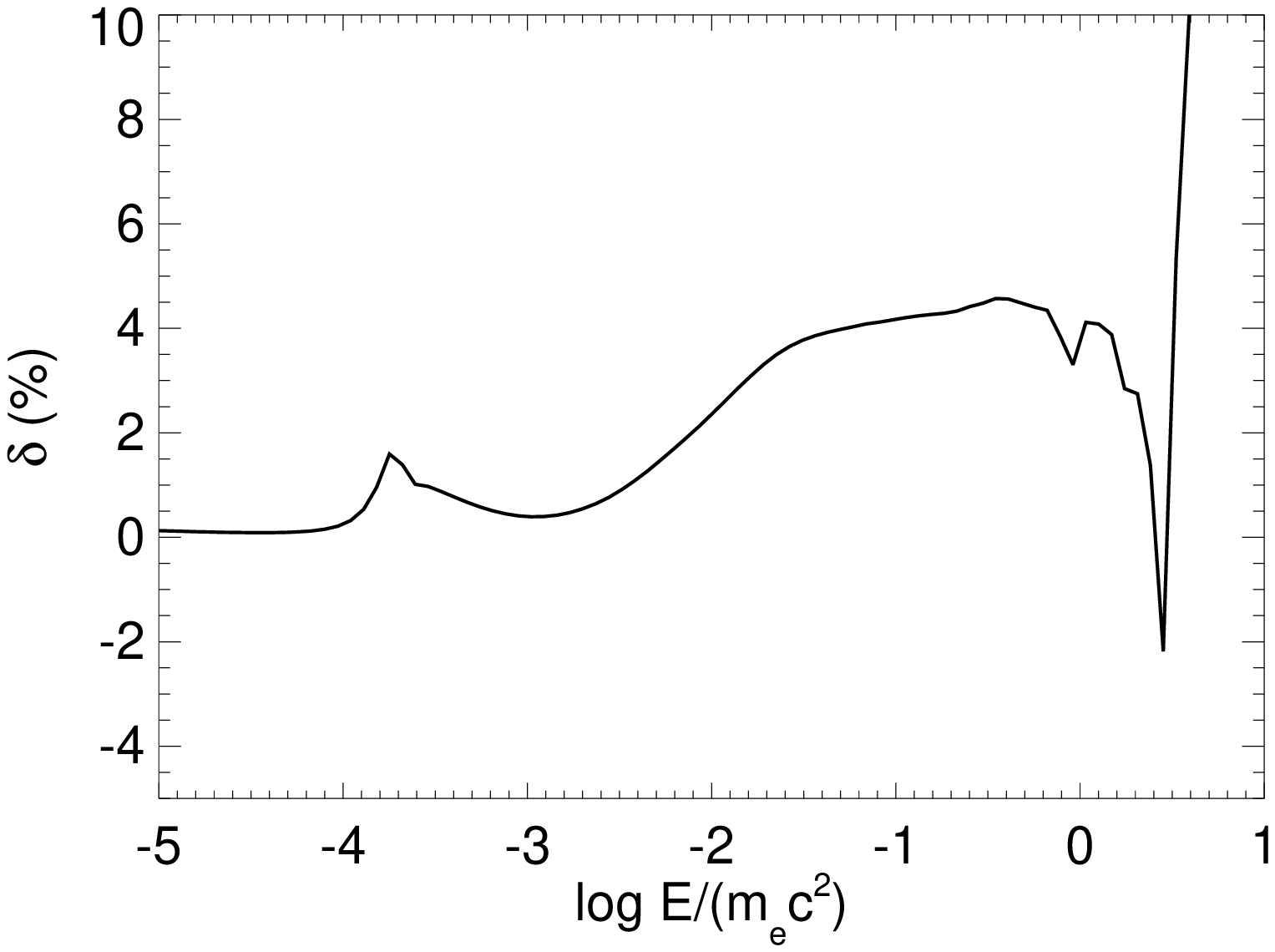}}
\end{center}
\end{figure}

In Figure \ref{fig:poutanen1} the
corona temperature is 56 keV with optical depth $\tau=0.5$, and in
Figure \ref{fig:poutanen2} $T_e=352$ keV and $\tau=0.05$,
corresponding to Figures 5 and 6 in \citet{poutanen:96}. In the upper
panels we show the observed flux at two inclination angles $\cos
i=0.11,0.5$, and in the bottom panels we show the polarization degree
$\delta(\%)=Q/I \times 100$. In all panels, the solid curves correspond to the
total flux, while the (dotted, dashed, dot-dashed, triple-dot-dashed,
and long-dashed) curves represent subsets of the flux, binned by
number of coronal scatterings (0, 1, 2, 3, $\ge 5$). Photon packets
that return to the disk suffer photoelectric absorption and Compton
recoil losses, and are then launched again from the disk, resetting
$N_{\rm scat}$ to zero. Thus the dotted curves in Figures
\ref{fig:poutanen1} and \ref{fig:poutanen2} have significant power
around the Compton hump at 10-100 keV. As discussed in
\citet{schnittman:10}, more scatterings in a sandwich corona
effectively constrain the geometry and increase the amplitude of
polarization at high energies.

We find excellent agreement overall, but are clearly dominated by
Monte Carlo noise above $\sim 100$ keV. For these disk and coronal
parameters, this corresponds to seed photons that have already
scattered on average over 25 times, so it is very difficult to
resolve any polarization signal at the few percent
level. Additionally, due to our photon packet algorithm, we are
limited to energy-independent electron cross sections, so we should
expect that the accuracy of our spectral predictions breaks down much
above 100 keV anyway. 

\section{CONCLUSION}\label{section:conclusion}

We have presented the technical details behind the general
relativistic radiation transport code \pand. Its capabilities include
optically thin emission and absorption, Compton scattering,
polarization, spectral and timing analysis, and flexible geometries
that allow analysis of numerous accretion models and MHD
simulations. We have discussed a number of practical challenges that
may also face other teams working to develop similar ray-tracing
codes, such as the method of weights in the scattering kernel. 

This is by no means the final word on \pand. Its great strength lies
in its flexibility, and we envisage numerous upgrades and improvements
in the near future. These will include, but not be limited to,
detailed ionization balance in the disk photosphere for improved AGN
modeling, time interpolation between simulation snapshots for
generating more accurate light curves, and the inclusion of
more sophisticated emission and absorption processes (e.g.,
angle-dependent synchrotron) to model low-luminosity sources such as Sgr
A$^\ast$. Perhaps most important, we will work to close the final
remaining gap between theory and observation by incorporating \pan
spectra into a data analysis framework like {\tt xspec} and making it
publicly available to the X-ray astronomy community.

\appendix

\section{Hamiltonian Equations of Motion}\label{section:app_ham}
The equations of motion for the Hamiltonian $H$ in
Boyer-Lindquist coordinates, as given in Section
\ref{section:raytracing}, are repeated here for completeness:
\begin{displaymath}
H(r,\theta,\phi,p_r,p_\theta,p_\phi) = -p_t = \omega p_\phi
+\alpha\left(\frac{\Delta}{\rho^2}p_r^2 
+\frac{1}{\rho^2}p_\theta^2 +\frac{1}{\varpi^2} p_\phi^2
+m^2\right)^{1/2},
\end{displaymath}
and according to classical theory:
\begin{subequations}
\begin{eqnarray}
\frac{dx^i}{dt} &=& \frac{\partial H}{\partial p_i} \, ,\\
\frac{dp_i}{dt} &=& -\frac{\partial H}{\partial x^i} \, .
\end{eqnarray}
\end{subequations}
For convenience of notation, we define the quantity $D^2$ as
\begin{equation}\label{def_X}
D^2(r,\theta,\phi,p_r,p_\theta,p_\phi) =
\frac{\Delta}{\rho^2}p_r^2 
+\frac{1}{\rho^2}p_\theta^2 +\frac{1}{\varpi^2} p_\phi^2
+m^2 \, .
\end{equation}
Then for an arbitrary variable $y \in (x^i,p_i)$, the partial derivative
of $H$ can be written
\begin{equation}\label{dH_dy}
\frac{\partial H}{\partial y} = \frac{\partial}{\partial y} 
(\omega p_\phi) + \frac{\partial \alpha}{\partial y}D -
\frac{1}{2}\frac{\alpha^2}{p_t+\omega p_\phi} 
\frac{\partial D^2}{\partial y} \, .
\end{equation}
The first set of Hamiltonian's equations are straightforward to
produce: 
\begin{subequations}
\begin{eqnarray}\label{dxi_dt}
\frac{dr}{dt} &=& \frac{\partial H_1}{\partial p_r} =
-\frac{p_r}{p_t+\omega p_\phi}\frac{\alpha^2 \Delta}{\rho^2} \, , \\
\frac{d\theta}{dt} &=& \frac{\partial H_1}{\partial p_\theta} =
-\frac{p_\theta}{p_t+\omega p_\phi}\frac{\alpha^2}{\rho^2} \, , \\
\frac{d\phi}{dt} &=& \frac{\partial H_1}{\partial p_\phi} = \omega
-\frac{p_\phi}{p_t+\omega p_\phi}\frac{\alpha^2}{\varpi^2} \, .
\end{eqnarray}
\end{subequations}
The momentum equations are a bit more involved, but there are only two
of them (for $p_r$ and $p_\theta$; $p_\phi$ is conserved):
\begin{subequations}
\begin{eqnarray}\label{dpi_dt}
\frac{dp_r}{dt} =
-\frac{\partial \omega}{\partial r}p_\phi + \frac{p_t + \omega
 p_\phi}{\alpha} \frac{\partial \alpha}{\partial r} +
\frac{\alpha^2}{2(p_t + \omega p_\phi)}
 \left[\frac{\partial}{\partial r}
\left(\frac{\Delta}{\rho^2}p_r^2 +\frac{1}{\rho^2}p_\theta^2 
+\frac{1}{\varpi^2} p_\phi^2 \right)\right] \, , \\
\frac{dp_\theta}{dt} =
-\frac{\partial \omega}{\partial \theta}p_\phi + \frac{p_t + \omega
 p_\phi}{\alpha} \frac{\partial \alpha}{\partial \theta} +
\frac{\alpha^2}{2(p_t + \omega p_\phi)}
 \left[\frac{\partial}{\partial \theta}
\left(\frac{\Delta}{\rho^2}p_r^2 +\frac{1}{\rho^2}p_\theta^2 
+\frac{1}{\varpi^2} p_\phi^2 \right)\right] \, .
\end{eqnarray}
\end{subequations}
The relevant spatial derivatives are as follows:
\begin{subequations}
\begin{eqnarray}
\frac{\partial\omega}{\partial r} &=& -\frac{\omega^2}{2Ma}
\left[3r^2+a^2(1+\cos^2\theta)-\frac{a^4}{r^2}\cos^2\theta\right] \, , \\
\frac{\partial\omega}{\partial \theta} &=& -\frac{\omega^2}{2Ma}
\left[\left(2Ma^2-a^2r-\frac{a^4}{r}\right)\sin\theta\cos\theta
  \right] \, , \\
\frac{\partial\alpha}{\partial r} &=& \frac{1}{2\alpha} \frac{\partial
  \alpha^2}{\partial r} \, , \\
\frac{\partial\alpha}{\partial \theta} &=& \frac{1}{2\alpha} \frac{\partial
  \alpha^2}{\partial \theta} \, , \\
\frac{\partial\alpha^2}{\partial r} &=& -\alpha^4
\left(\frac{2M}{\Delta\rho^2}\right)
\left(\frac{a^4-r^4}{\Delta}-\frac{2r^2a^2\sin^2\theta}{\rho^2}\right) \, , \\
\frac{\partial\alpha^2}{\partial \theta} &=& -\alpha^4
\left[\frac{4Ma^2r\sin\theta\cos\theta(a^2+r^2)}{\Delta\rho^2}\right] \, , \\
\frac{\partial}{\partial r}\left(\frac{1}{\varpi^2}\right) &=& 
-\frac{2}{\varpi^4}\left[\sin^2\theta\left(r+\frac{2Ma^2\sin^2\theta
(a^2\cos^2\theta-r^2)}{\rho^4}\right)\right] \, , \\
\frac{\partial}{\partial \theta}\left(\frac{1}{\varpi^2}\right) &=& 
-\frac{4\sin\theta\cos\theta}{\varpi^4}\left[2Ma^2\sin^2\theta
\left(\frac{r^2+a^2}{\rho^4}+\frac{1}{\rho^2}\right)+(r^2+a^2)\right] \, , \\
\frac{\partial}{\partial r}\left(\frac{\Delta}{\rho^2}\right) &=& 
\frac{2}{\rho^2}\left(r-M-\frac{r\Delta}{\rho^2}\right) \, , \\
\frac{\partial}{\partial \theta}\left(\frac{\Delta}{\rho^2}\right) &=& 
\frac{2}{\rho^4}a^2\Delta\sin\theta\cos\theta \, , \\
\frac{\partial}{\partial r}\left(\frac{1}{\rho^2}\right) &=& 
-\frac{2r}{\rho^4} \, , \\
\frac{\partial}{\partial \theta}\left(\frac{1}{\rho^2}\right) &=& 
\frac{2}{\rho^4}a^2\sin\theta\cos\theta \, .
\end{eqnarray}
\end{subequations}

\section{Monte Carlo Sampling of Maxwell-Juttner
Distribution}\label{section:app_juttner} 

For any normalized distribution function $f(x)$ with $x \in
(-\infty,\infty)$, one can always define the cumulative distribution
function 
\begin{equation}
{\rm cdf}(x) = F(x) =\int_{-\infty}^x f(x')dx' \, ,
\end{equation}
with $F(-\infty)=0$ and $F(\infty)=1$. Then by selecting a uniform
random number $\lambda \in [0,1)$, the choice $x=F^{-1}(\lambda)$ will
be distributed according to $f(x)$. However, in most cases, $F(x)$
cannot be written in closed form, so other methods are required. 

One simple technique described in \citet{press:92} is the ``rejection
method,'' where an auxiliary function $g(x)$ is used, where
$g(x)>f(x)$ everywhere, and $G(x)$ is easy to calculate. We begin by
selecting a trial $x_0=G^{-1}(\lambda_0)$, then pick another random
deviate $\lambda_1$. If $\lambda_1 < f(x_0)/g(x_0)$ then $x_0$ is
selected as a representative sample of $f(x)$, else we try again with
a new $\lambda_0$. Of course, if $g(x)$ is large enough, it is easy to
ensure that it is greater than $f(x)$ everywhere. However, the
efficiency of this method is limited by the ratio of the areas under
the two curves $f(x)$ and $g(x)$, so it is desirable to pick $g(x)$ as
close to $f(x)$ as possible \citep{press:92}.

For the Maxwell-Juttner distribution defined in equation
(\ref{eqn:maxwell_juttner}):
\begin{equation}
f(\gamma) \sim \gamma^2\beta \exp(-\gamma/\theta_T)\, ,
\end{equation}
we choose an auxiliary function
\begin{equation}
g(\gamma) \sim \gamma^2 \exp(-\gamma/\theta_T)\, .
\end{equation}
This gives
\begin{equation}\label{eqn:G_cdf}
G(\gamma) = 1-\frac{e^{-\gamma/\theta_T}}{e^{-1/\theta_T}}
\frac{2\theta_T^2+2\theta_T \gamma+\gamma^2}{2\theta_T^2+2\theta_T+1}
\end{equation}
for the cumulative distribution function. Inverting (\ref{eqn:G_cdf})
isn't trivial, but can be done numerically with a simple root
finder. For these choices of $f(\gamma)$ and $g(\gamma)$, we find
excellent efficiency for this algorithm of $\sim 90$\%. 

\section{Comparison of Scattering Kernels}\label{section:app_kernel}
As described in Section \ref{section:scattering}, there are (at least)
two different ways to implement the scattering of polarized light off
of free electrons. 

The method of weights picks a random scattering angle from a uniform
distribution of $\cos\theta \in [-1,1]$ and $\phi \in [0,2\pi)$, then
weights the scattered beam of photons by the cross section in that
direction, normalized by the average cross section to conserve
flux. By integrating equation (\ref{eqn:cross_general}) over $\phi$,
this resembles the classical cross section for unpolarized light:
\begin{equation}\label{eqn:w_thomson}
w(\theta) = \frac{I'}{I}=\frac{3}{4}(\cos^2\theta+1).
\end{equation}

Because repeated scatters tend to increase the level of polarization
(indeed, in the microscopic limit, {\it every} photon has $\delta=1$),
we will focus on the case where $\delta=1$, giving
\begin{equation}
w(\theta,\phi)=\frac{3}{2}(\cos^2\theta \cos^2\phi+\sin^2\phi).
\end{equation}
For angles uniformly distributed in $\cos\theta$ and $\phi$, one can
show that the probability distribution
function (pdf) for $w$ is
\begin{equation}\label{eqn:P_w1}
P(w) = \frac{1}{3}\left(1-\frac{2}{3}w\right)^{-1/2}
\end{equation}
for $0\le w \le 3/2$, and 0 otherwise.

For multiply-scattered photons, the weight function is multiplicative,
since the individual scattering events are uncorrelated. For $n$
scatters, the net weight is given by
\begin{equation}
W = \prod_{i=1}^{n}w_i.
\end{equation}
To determine the pdf $P(W)$, we define a new variable $Z$:
\begin{equation}
Z \equiv \ln W = \sum_{i=1}^{n} \ln w_i = \sum_{i=1}^{n} z_i .
\end{equation}
For large values of $n$, the central limit theorem dictates that the
distribution of $Z$ should be Gaussian:
\begin{equation}\label{eqn:P_Z}
P(Z) = \frac{1}{\sigma_z\sqrt{2\pi n}}\exp
\left(-\frac{(Z-n\mu_z)^2}{2n\sigma_z^2}\right),
\end{equation}
where $\mu_z$ and $\sigma_z^2$ are respectively the mean and variance
of $P(z)$. From equation (\ref{eqn:P_w1}) and the variable transformation
$z=\ln w$, we have 
\begin{equation}
P(z) =
P(w)\frac{dw}{dz}=\frac{e^z}{3}\left(1-\frac{2}{3}e^z\right)^{-1/2}, 
\end{equation}
with $z \in (-\infty,\ln 3/2]$. This gives $\mu_z=-0.208$ and
$\sigma_z^2=0.710$. Now we see that the pdf $P(W)$ is given by a
log-normal distribution: 
\begin{equation}
P(W) = \frac{1}{W\sigma_z\sqrt{2\pi n}} 
\exp\left(-\frac{(\ln W-n\mu_z)^2}{2n\sigma_z^2}\right).
\end{equation}

For photons random-walking through an atmosphere of optical depth $\tau$,
we find the pdf of the number of scatters required to escape can be
closely approximated by
\begin{equation}
P(n) = \frac{n}{4\tau^2} \exp\left(-\frac{n}{2\tau}\right) .
\end{equation}
Then the net distribution $P(W)$ for all scatting orders is simply
\begin{equation}
P(W;\tau) = \int_0^\infty dn\, P(n;\tau)\, P(W;n).
\end{equation}
 
\begin{figure}
\caption{\label{fig:PW2} Relative contribution to the total observed
  spectrum by photons of a given weight, for optical depths of
  $\tau=(1,2,3,5,10)$. Any calculation with $\tau\gtrsim2$ will not formally
  converge using the method of weights.}
\begin{center}
\includegraphics[width=0.6\textwidth]{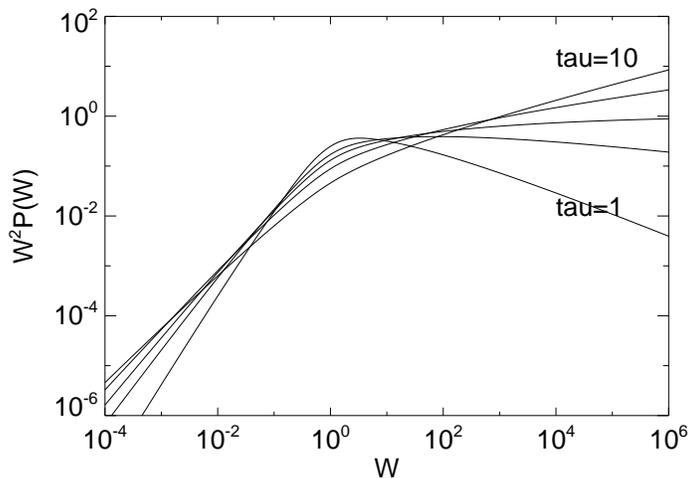}
\end{center}
\end{figure}

The relative contribution to the spectrum from photons with a weight
in the range
$(W, W+dW)$ is $P(W)WdW$, so we require $P(W)$ to decrease faster than
$W^{-2}$ for large $W$ if the calculation is to converge. In Figure
\ref{fig:PW2} we plot $W^2 P(W)$ for a range of
$\tau$. Our analytic results suggest that for $\tau \gtrsim 2$, any
polarization spectrum formed using
this Monte Carlo weighting method should be dominated by the rarest,
highest-weight photon packets, confirming what we have seen in
trial runs with large $\tau$. Now, in practice, the convergence
is not quite as bad as Figure \ref{fig:PW2} suggests, for two primary
reasons. First, the seed photon packets have little or no
polarization, so the initial weighting function more closely resembles
equation (\ref{eqn:w_thomson}), which leads to a significantly tighter
range in $W$: $\mu_{\rm unpol}=-0.027$ and $\sigma_{\rm unpol}^2 =
0.047$ (in this unpolarized limit, the weight method converges for all
optical depths up to $\tau \gtrsim 200$). 
Second, for the small-to-moderate optical depths of $\tau
\lesssim 5$, the typical number of scatters $n$ is still small enough
that the mean value theorem does not strictly apply, in effect cutting
off the high-weight tails in (\ref{eqn:P_Z}) and further reducing the
contribution from statistical outliers. 

\begin{figure}
\caption{\label{fig:PW2MC} Relative contribution to the total observed
  spectrum by photons of a given weight, for optical depths of
  $\tau=(2,5,10)$, sorted by color (red, blue, black). The solid lines
  are the analytic results, and the crosses are ``data'' from 
  Monte Carlo calculations of $10^6$ photons each.}
\begin{center}
\includegraphics[width=0.6\textwidth]{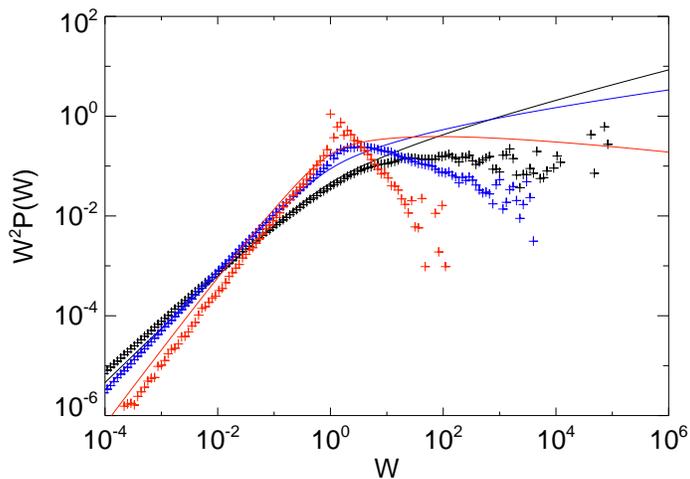}
\end{center}
\end{figure}

However, for $\tau \gtrsim 10$, the polarization of a typical photon
bundle reaches $\delta \to 1$
after just a few scatters, and the large number of total scattering
events allows us to reproduce these analytic results with numerical
tests of the Monte Carlo code. In Figure \ref{fig:PW2MC} we show the
distribution of weights from a calculation using unpolarized seed photons,
scattering through optical depths of $\tau=2,5,10$. While we find that
the $\tau=5$ case does converge eventually, in practice we find the
convergence is so slow that another Monte Carlo method is
preferrable. Furthermore, the highest weights have the fewest events,
and thus also suffer from small-number statistics, potentially adding
to the ``undue influence'' of outliers. This can be seen in the
scatter at the high-weight end of each data set.

Instead of picking a scattering angle at random and weighting it by
equation (\ref{eqn:I_Iprime}), let us use the differential
cross section (\ref{eqn:cross_general}) to get the scattering pdf:
\begin{equation}\label{eqn:P_thetapsi}
P(\theta,\phi) = \frac{3}{16\pi} [(1-\delta)(\cos^2\theta+1) +
  2\delta\cos^2\theta \cos^2\phi + 2\delta\sin^2\phi]\, .
\end{equation}
Integrating over $\phi$, we again find the standard Thomson cross
section (this holds even for $\delta \ne 0$):
\begin{equation}
P(\cos\theta) = \frac{3}{8}(\cos^2\theta+1).
\end{equation}
Writing $\mu\equiv\cos\theta$ for convenience, the cumulative
distribution function is given by 
\begin{equation}\label{eqn:cdf_theta}
{\rm cdf}(\mu) = \int_{-1}^{\mu} P(\mu')d\mu'=\frac{1}{8}(\mu^3+3\mu+4)
\end{equation}
To pick an appropriate value for $\mu$, generate a random number
uniformly distributed $\lambda \in [0,1)$, and invert equation
(\ref{eqn:cdf_theta}), in effect solving for the root of the cubic:
\begin{equation}
\mu^3 + 3\mu+4-8\lambda=0.
\end{equation}
Because ${\rm cdf}(\mu)$ is monotonically increasing, this
equation is guaranteed to have a single real root in the interval $-1
\le \mu \le 1$. 

Once $\mu$ is selected, we choose $\phi$ by the same method,
now using the pdf 
\begin{equation}
P(\phi; \mu) = \frac{1}{2\pi(\mu^2+1)} [(1-\delta)(\mu^2+1) +
  2\delta\mu^2 \cos^2\phi + 2\delta\sin^2\phi]\, ,
\end{equation}
which gives
\begin{equation}\label{eqn:cdf_phi}
{\rm cdf}(\phi; \mu) = \frac{\phi}{2\pi}+
\frac{\delta}{4\pi}\left(\frac{\mu^2-1}{\mu^2+1}\right)\sin(2\phi).
\end{equation}
Again, to pick an appropriate $\phi$ given a uniform random $\lambda$,
one must invert equation (\ref{eqn:cdf_phi}) to get $\phi={\rm
  cdf}^{-1}(\lambda)$. Unfortunately, this is equivalent to solving
Kepler's equation, which has no closed-form solution, and must be done
numerically. Fortunately, this is equivalent to solving Kepler's
equation, one of the best-studied numerical problems in astrophysics!
In practice, we use the iterative approach outlined in
\citet{murray:99}. While slightly more time consuming than the method
of weights, the exact cross section method has the distinct advantage
of converging for an arbitrary number of scatterings, and thus is the
method we prefer for \pand.

\newpage


\newpage
\begin{thebibliography}{99}
\bibitem[Agol \& Krolik(2000)]{agol:00} Agol, E., \& Krolik, J.\
  H. 2000, ApJ, 528, 161
\bibitem[Bardeen et al.(1972)]{bardeen:72} Bardeen, J.\ M., Press,
  W.\ H., \& Teukolsky, S.\ A. 1972, ApJ 178, 347
\bibitem[Beckwith et al.(2008)]{beckwith:08} Beckwith, K., Hawley, J.F.
  \& Krolik, J.H. 2008, MNRAS 390, 21
\bibitem[Boyer \& Lindquist(1967)]{boyer:67} Boyer, R.\ H., \&
  Lindquist, R.\ W. 1967, J.\ Math.\ Phys., 8, 265
\bibitem[Broderick \& Blandford(2003)]{broderick:03} Broderick, A.\
  E., \& Blandford, R. 2003, MNRAS, 342, 1280 
\bibitem[Broderick \& Blandford(2004)]{broderick:04} Broderick, A.\
  E., \& Blandford, R. 2004, MNRAS, 349, 994 
\bibitem[Carter(1968)]{carter:68} Carter, B. 1968, Phys.\ Rev.\ Lett.\
  26, 331
\bibitem[Chandrasekhar(1960)]{chandra:60} Chandrasekhar, S.
  1960. {\it Radiative Transfer}, Dover, New York
\bibitem[Connors \& Stark(1977)]{connors:77} Connors, P.\ A., \&
  Stark, R.\ F. 1980, Nature, 269, 128
\bibitem[Connors et al.(1980)]{connors:80} Connors, P.\ A., Piran,
  T., \& Stark, R.\ F. 1980, ApJ, 235, 224
\bibitem[Dexter \& Agol(2009)]{dexter:09} Dexter, J., \& Agol,
  E. 2009, ApJ, 696, 1616
\bibitem[Dexter et al.(2009)]{dexter:09b} Dexter, J., Agol, E., \&
  Fragile, P.C. 2009, ApJ, 703, L142 
\bibitem[Dexter et al.(2010)]{dexter:10} Dexter, J., Agol, E.,
  Fragile, P.C., \& McKinney, J.C. 2010, ApJ, 717, 1092
\bibitem[Dexter et al.(2012)]{dexter:12} Dexter, J., McKinney, J.C., \&
  Agol, E. 2012, MNRAS, 421, 1517 
\bibitem[Dolence et al.(2009)]{dolence:09} Dolence, J.\ C., Gammie,
  C.\ F., Moscibrodzka, M., \& Leung, P.\ K. 2009, ApJS, 184, 387
\bibitem[Dovciak et al.(2004)]{dovciak:04} Dovciak, M., Karas, V., \&
  Yaqoob, T. 2004, ApJS, 153, 205
\bibitem[Dovciak et al.(2008)]{dovciak:08} Dovciak, M., Muleri, F.,
  Goosmann, R.\ W., Karas, V., \& Matt, G. 2008, MNRAS, 391, 32
\bibitem[Dovciak et al.(2011)]{dovciak:11} Dovciak, M., Muleri, F.,
  Goosmann, R.\ W., Karas, V., \& Matt, G. 2011, ApJ, 731, 75
\bibitem[Haardt \& Maraschi(1993)]{haardt:93} Haardt, F., \& Maraschi,
  L. 1993, ApJ, 413, 507
\bibitem[Haardt et al.(1994)]{haardt:94} Haardt, F., Maraschi,
  L., \& Ghisellini, G. 1994, ApJ, 432, L95
\bibitem[Huang et al.(2009)]{huang:09} Huang, L., Liu, S., Shen,
  Z.-Q., Yuan, Y.-F., Cai, M.\ J., Li, H., \& Fryer, C.\ L. 2009, ApJ,
  703, 557 
\bibitem[Huang \& Shcherbakov(2011)]{huang:11} Huang, L., \&
  Shcherbakov, R.\ V. 2011, MNRAS, 416, 2574 
\bibitem[Johannsen \& Psaltis(2010a)]{johannsen:10a} Johannsen, T., \&
  Psaltis, D. 2010a, ApJ, 716, 187
\bibitem[Johannsen \& Psaltis(2010b)]{johannsen:10b} Johannsen, T., \&
  Psaltis, D. 2010b, ApJ, 718, 446
\bibitem[Johannsen \& Psaltis(2011)]{johannsen:11} Johannsen, T., \&
  Psaltis, D. 2011, ApJ, 726, 11
\bibitem[Johannsen \& Psaltis(2012)]{johannsen:12} Johannsen, T., \&
  Psaltis, D. 2012, ApJ submitted [arXiv:1202.6069]
\bibitem[Kojima(1991)]{kojima:91} Kojima, Y. 1991, MNRAS, 250, 629
\bibitem[Krawczynski(2012)]{krawczynski:12} Krawczynski, H. 2012,
  ApJ, 754, 133 
\bibitem[Laor et al.(1990)]{laor:90} Laor, A., Netzer, H., \& Piran,
  T. 1990, MNRAS, 242, 560
\bibitem[Laor(1991)]{laor:91} Laor, A. 1991, ApJ, 376, 90
\bibitem[Li et al.(2008)]{li:08} Li, L.-X., Narayan, R., \&
  McClintock, J.\ E. 2008, ApJ submitted, [arXiv:0809.0866]
\bibitem[Marin et al.(2012)]{marin:12} 	Marin, F., Goosmann, R. W.,
  Dovciak, M., Muleri, F., Porquet, D., Grosso, N., Karas, V., \&
  Matt, G. 2012, MNRAS, 426, L101
\bibitem[Matt et al.(1993)]{matt:93} Matt, G., Fabian, A.\ C., \&
  Ross, R.\ R. 1993, MNRAS, 264, 839
\bibitem[Misner et al.(1973)]{mtw:73} Misner, C.\ W., Thorne, K.\ S.,
  \& Wheeler, J.\ A. 1973, {\it Gravitation} (W.\ H.\ Freeman, San
  Francisco)
\bibitem[Morrison \& McCammon(1983)]{morrison:83} Morrison, R., \&
  McCammon, D. 1983, ApJ, 270, 119
\bibitem[Murray \& Dermott(1999)]{murray:99} Murray, C.\ D., \&
  Dermott, S.\ F. 1999, {\it Solar System Dynamics}, Cambridge
  University Press, Cambridge
\bibitem[Noble et al.(2007)]{noble:07} Noble, S.\ C., Leung, P.\ K.,
  Gammie, C.\ F., Book, L.\ G. 2007, CQG, 24, S259
\bibitem[Noble et al.(2009)]{noble:09} Noble, S.\ C., Krolik, J.\ H.,
  \& Hawley, J.\ F. 2009, ApJ, 692, 411
\bibitem[Noble et al.(2010)]{noble:10} Noble, S.\ C., Krolik, J.\ H.,
  \& Hawley, J.\ F. 2010, ApJ 711, 959
\bibitem[Noble et al.(2011)]{noble:11} Noble, S.\ C., Krolik, J.\ H.,
  Schnittman, J.\ S., \& Hawley, J.\ F. 2011, ApJ, 743, 115
\bibitem[Novikov \& Thorne(1973)]{novikov:73} Novikov, I.\ D., \& Thorne,
  K.\ S. 1973, in \textit{Black Holes}, ed. C.\ DeWitt \& B.\ S.\
  DeWitt (New York: Gordon and Breach)
\bibitem[Poutanen \& Svensson(1996)]{poutanen:96} Poutanen, J., \&
  Svensson, R. 1996, ApJ, 470, 249
\bibitem[Rauch \& Blandford(1994)]{rauch:94} Rauch, K.\ P., \& Blandford,
  R.\ D. 1994, ApJ, 421, 46
\bibitem[Press et al.(1992)]{press:92} Press, W.\ H., et al. 1992,
  \textit{Numerical Recipes in C: The Art of Scientific Computing}
  (Cambridge: Cambridge University Press)
\bibitem[Psaltis \& Johannsen(2012)]{psaltis:12} Psaltis, D., \&
  Johannsen, T. 2012, ApJ, 745, 1
\bibitem[Rybicki \& Lightman(2004)]{rybicki:04} Rybicki, G.\ B., \&
  Lightman, A.\ P. 2004, \textit{Radiative Processes in Astrophysics}
  (Weinheim: Wiley-VCH)
\bibitem[Schnittman \& Craxton(1996)]{schnittman:96} Schnittman, J.\
  D., \& Craxton, R.\ S. 1996, Phys.\ Plasmas, 3, 3786
\bibitem[Schnittman \& Craxton(2000)]{schnittman:00} Schnittman, J.\
  D., \& Craxton, R.\ S. 2000, Phys.\ Plasmas, 7, 2964

\bibitem[Schnittman \& Bertschinger(2004)]{schnittman:04} Schnittman,
  J.\ D., \& Bertschinger, E. 2004, ApJ, 606, 1098
\bibitem[Schnittman \& Rezzolla(2006)]{schnittman:06a} Schnittman, J.\
  D., \& Rezzolla, L. 2006, ApJ, 637, L113
\bibitem[Schnittman et al.(2006)]{schnittman:06} Schnittman, J.\
  D., Krolik, J.\ H., \& Hawley, J.\ F. 2006, ApJ, 651, 1031
\bibitem[Schnittman \& Krolik(2009)]{schnittman:09} Schnittman, J.\
  D., \& Krolik, J.\ H. 2009, ApJ 701, 1175
\bibitem[Schnittman \& Krolik(2010)]{schnittman:10} Schnittman, J.\
  D., \& Krolik, J.\ H. 2010, ApJ, 712, 908
\bibitem[Schnittman et al.(2012)]{schnittman:12} Schnittman, J.\
  D., Krolik, J.\ H., \& Noble, S.\ C. 2012, ApJ submitted,
  arXiv:1207.2693 
\bibitem[Shcherbakov \& Huang(2011)]{shcherbakov:11} Shcherbakov, R.\
  V., \& Huang, L. 2011, MNRAS, 410, 1052 
\bibitem[Shimura \& Takahara(1995)]{shimura:95} Shimura, T., \&
  Takahara, F. 1995, ApJ, 445, 780
\bibitem[Sunyaev \& Titarchuk(1985)]{sunyaev:85} Sunyaev, R.\ A., \&
  Titarchuk, L.\ G. 1985, A\&A, 143, 374
\bibitem[Walker \& Penrose(1970)]{walker:70} Walker, M., \& Penrose,
  R. 1970, Commun.\ Math.\ Phys., 18, 265
\end{thebibliography}
\end{document}